\documentclass[twocolumns]{aa}  

\usepackage{graphicx}
\usepackage{subfig}
\usepackage{enumitem}
\usepackage{txfonts}
\usepackage{hyperref}
\usepackage{mathtools}
\usepackage{xcolor}
\usepackage{ulem}

\def\ls{{_<\atop^{\sim}}}
\def\gs{{_>\atop^{\sim}}}

\begin{document} 

\title{Time Evolving Photo Ionisation Device (TEPID): a novel code for out-of-equilibrium gas ionisation}
\author{
A. Luminari\inst{1,2} \and F. Nicastro\inst{2} \and Y. Krongold\inst{3} \and L. Piro\inst{1} \and A. L. Thakur\inst{1,4}
}

\institute{
INAF - Istituto di Astrofisica e Planetologia Spaziali, Via del Fosso del Caveliere 100, I-00133 Roma, Italy
\and
INAF - Osservatorio Astronomico di Roma, Via Frascati 33, 00078 Monteporzio, Italy
\and
Instituto de Astronomía, Universidad Nacional Autónoma de México, Circuito Exterior, Ciudad Universitaria, Ciudad de México 04510, México
\and
Dipartimento di Fisica, Universit\`a degli Studi di Roma "Tor Vergata", Via della Ricerca Scientifica 1, 00133 Roma RM, Italy
}
\date{Received 1 December 2022; accepted 29 August 2023}

\abstract
{Photoionisation is one of the main mechanisms at work in the gaseous environment of bright astrophysical sources. Many information on the gas physics, chemistry and kinematics, as well as on the ionising source itself, can be gathered through optical to X-ray spectroscopy. While several public time equilibrium photoionisation codes are readily available and can be used to infer average gas properties at equilibrium, time-evolving photoionisation models have only very recently started to become available. They are needed when the ionising source varies faster than the typical gas equilibration timescale. Indeed, using equilibrium models to analyse spectra of non-equilibrium photoionised gas may lead to inaccurate results and prevents a solid assessment of the gas density, physics and geometry.}
{Our main objective is to present and make available the Time-Evolving PhotoIonisation Device (TEPID), a new code that self-consistently solves time evolving photoionisation equations (both thermal and ionisation balance) and accurately follows the response of the gas to changes of the ionising source.}
{TEPID self-consistently follows the gas temperature and ionisation in time by including all the main ionisation/recombination and heating/cooling mechanisms.
The code takes in input the ionising lightcurve and spectral energy distribution and solves the time-evolving equations as a function of the gas electron density and of the time. The running time is intelligently optimised by an internal algorithm that initially scans the input lightcurve to set a time-dependent integration frequency. The code is built in a modular way, can be applied to a variety of astrophysical scenarios and produces time-resolved gas absorption spectra to fit the data.} 
{To describe the structure and main features of the code we present two applications of TEPID to two dramatically different astrophysical scenarios: the typical ionised absorbers observed in the X-ray spectra of Active Galactic Nuclei (e.g. Warm Absorbers and UFOs), and the circumburst environment of a Gamma-Ray Burst. For both cases, we show how the gas energy and ionisation balances vary as a function of time, gas density and distance from the ionising source. We show that time evolving photoionisation leads to unique ionisation patterns which cannot be reproduced by stationary photoionisation codes when the gas is out of equilibrium. This demonstrates the need for codes such as TEPID in view of the  unprecedented capabilities that will be offered by the up-coming high-resolution X-ray spectrometers onboard missions like XRISM or Athena.}
{}


\titlerunning{The TEPID Time Evolving Code}
\authorrunning{A. Luminari et al.}
\maketitle

\section{Introduction}
Photoionisation is an ubiquitous process in astrophysics. Any gaseous astrophysical environment embedded in a field of ionising photons (E>13.6 eV for ground state Hydrogen) gets photoionised to a level that depends critically on the ratio between the density of ionising photons at the illuminated face of the cloud and the electron density in the gas, i.e., the ionisation parameter $U=(Q_{ion}/4 \pi r^2 c)/n_H$, where $Q_{ion}$ is the rate of ionising photons emitted by the source and $n_H,r,c$ are the hydrogen electron density, the distance of the illuminated face of the gas cloud from the ionising source and the speed of light, respectively \footnote{A widely-used alternative definition for the ionisation parameter is $\xi=L_{ion}/n_e^{tot} r^2$, where $L_{ion}, n_e^{tot}$ are the ionising luminosity and the total electron density \citep{tarter69}}. A few important examples are the gaseous environment of active galactic nuclei (AGNs) and accreting compact sources, the interstellar-medium (ISM) surrounding explosive events like Gamma-Ray Bursts (GRBs), the intergalactic medium embedded in the diffuse radiation field and the HII regions around O and B stars. 

Diagnostics on the ionising source and on the physics, kinematics and chemistry of the ionised gas can be gathered through UV and X-ray spectroscopy, combined with gas ionisation modelling. 
Several public photoionisation codes do exist, which describe (given a set of input parameters) the stationary state of a gas cloud at equilibrium (e.g. Cloudy, \citealp{Cloudy17}; XSTAR, \citealp{xstar,kallman21}; SPEX, \citealp{spex}).
Albeit each one has its own working structure and atomic libraries, they generally model equilibrium configurations, i.e. assume gas in local photoionisation equilibrium. In such circumstances, for a given incident spectral energy distribution and metallicity, the ionic population of the gas, and thus the emerging absorption spectrum, is uniquely determined by the value of $U$ \footnote{We also note that for high densities, $n_H >> 10^{11} cm^{-3}$, stimulated processes can also affect the ionisation balance (see e.g. \citealp{kallman21}).}. From an observational measurement of $U, Q_{ion}$ it is possible to derive an estimate of the product $r^2 \cdot n_H$, but not independently of $n_H$ and $r$.

However, photoionisation equilibrium does not necessarily apply to gaseous environments illuminated by highly variable sources, such as those surrounding GRBs and AGNs, which are known to exhibit luminosity variations of up to several orders of magnitude.
In fact, when the ionising source varies on timescales shorter than the equilibration time-scale of the gas $t_{eq}$, equilibrium models can only provide an average description of the gas properties and may yield to the wrong diagnostic on the physics and ionisation state of the gas when compared to observations. The equilibration time $t_{eq}$ critically depends on the free electron density $n_e$ and (in a 3-ion approximation) can be expressed as \citep{nicastro99}:
\begin{equation}
t_{eq} \approx \Big\{ (\alpha_{rec}(n_{X^i},T) \cdot n_e) \cdot \Big(1+ \frac{\alpha_{rec}(n_{X^{i-1}},T)}{\alpha_{rec}(n_{X^i},T)}+\frac{n_{X^{i+1}}}{n_{X^i}} \Big) \Big\}^{-1}
\label{teq}
\end{equation}
where $\alpha_{rec},n_{X^i}$ are the recombination coefficient and the fractional abundance of the $i$-th ion of the element $X$, respectively, and $T$ is the electron temperature. We note that $t_{eq}$ accounts for the time required to reach a new equilibrium and, depending on whether the luminosity increases or decreases, is similar to the photoionisation and recombination timescales, respectively (see \citealp{garcia13,sadaula22}). In these cases, only time-evolving photoionisation models (combined with time-resolved spectroscopy) can provide instantaneous diagnostic on the physics and kinematics of the gas, and also offer an efficient way to remove the gas density-distance degeneracy intrinsic in the definition of $U$, both through changes in relative ion abundance (e.g. \citealp{nicastro99,kne07}) and electron population of metastable levels (e.g. \citealp{kaastra04}) in response to flux variations. This is the focus of the present paper.

GRB studies of the temporal variation of absorption lines in the optical/UV band have provided a wealth of information on the medium surrounding the burst and the ISM of the host galaxy along our line of sight, including its density, metallicity and distance from the burst (see e.g. \citealp{delia09,heintz18}). These results were based on ad-hoc time-evolving level-population models built under the UV photo-pumping hypothesis for a number of selected FeII transitions, rather than a general-purpose multi-wavelength time-evolving photoionisation code.
In the X-rays, the limited resolution and signal-to-noise ratio of current observations, together with the lack of self-consistent time-evolving photoionisation models, hampered so far detailed spectroscopic studies and investigations of the ionisation structure of the intervening gas. X-ray absorption in the $\sim 0.5-10$ keV portion of X-ray afterglow spectra has instead so far been modelled phenomenologically, adopting single-zone cold (i.e. neutral) photoelectric absorption models, which typically provide reasonable fits of the data (see e.g. \citealp{campana12,asquini19}).
Alternatively, single-zone, equilibrium ionised absorber modelling has been attempted on the X-ray spectra of X-ray bright and optically dark GRB afterglows (e.g. \citealp{Piro02}), which exhibit substantial absorption in X-rays and optical. These studies, however, provided only upper limits on the ionisation parameter, consistent with absorption by either dense absorbing clouds of the star-forming region embedding the GRBs or the more distant ISM of the host galaxy.

Time variability represents one of the best tools to explore the gaseous environment of AGNs as well. The response of nuclear absorbers to central source luminosity variations has provided fundamental insights into the gas distance, density, geometry and launching mechanisms in a number of sources, revealing a close interaction between the accretion-powered luminosity and the outflowing gas at accretion disc scales (e.g. \citealp{nicastro99,peterson04,kne07,bianchi09,ppf17,kara21}). These results, however, have been reached either through simple recombination and variability timescale comparisons or through simple time evolving photoionisation models (e.g. \citealp{nicastro99}).

Flexible, fast and self-consistent time-evolving photoionisation codes, able to produce time-resolved spectra for any kind of ionising spectral energy distribution (SED) and lightcurve over the whole rest-frame UV and X-ray bands, have only very recently started to become available \citep{sadaula22,rogantini22} and are urgently needed to fulfil the gap between current models and the new generation of upcoming spectrometers. 
In this paper we present the novel Time Evolving PhotoIonisation Device (TEPID). TEPID follows the energy and ion abundance evolution of a gas undergoing time variable photoionisation and, in its current version, computes the corresponding absorption spectra in the UV and X-ray bands. The main outputs of a TEPID run are the gas temperature, cooling and heating rates, ion abundances and column densities. Such quantities are fed to a customised version of the PHotoionised Absorber Spectral Engine (PHASE \citealp{krongold03}) which builds tables of time-resolved UV and X-ray absorption spectra for a number of input parameters (i.e. time, electron density, ion column density, etc.) that can be used in fitting packages (such as {\sc xspec}, \citealp{xspec}) to fit the data and constrain the properties of the absorbing gas. TEPID is highly flexible and can be applied for diagnostics to any of the astrophysical scenarios described above. It has been tested over a broad range of electron densities, equivalent hydrogen column densities and ionising fluxes (i.e. ionisation parameters) and compared, at equilibrium, with existing photoionisation codes. Main limitations of the current version of TEPID are: (1) time-evolving level population balance is not computed and all ions are considered in their ground states and (2) as a consequence, gas emissivity is not produced and only a first-order radial-opacity radiative-transfer is performed throughout the cloud (see \S \ref{discuss} for a discussion on these limitations).

In this first paper we present the code and its application to two different astrophysical cases: the highly ionised gaseous environment of AGNs and the post-explosion surroundings of GRBs. The two environments undergo dramatically different time-evolving photoionisations. In the first case, luminosity changes are stochastically distributed, both in time and magnitude, with up to few orders of magnitude intensity variations on timescales from $\sim$ ks to hundreds of ks and longer. In the second case, instead, the ionising light curve is typically described by an initial short high-luminosity phase (the so-called prompt that, for the purpose of this paper, can be approximated as constant), shortly followed by a steep and virtually monotonic decaying phase lasting several hundreds of ks, during which the luminosity drops by several orders of magnitudes (the afterglow).

The structure of the paper is as follows. In \S \ref{scheme} we describe the general structure of the code and its main outputs. Then, in \S \ref{agn-out} and \ref{grb-out} we present the results for the AGN and GRB cases, respectively. Finally, discussion and conclusions are in \S \ref{discuss} and \ref{conclus}.

\section{The TEPID code}
\label{scheme}
\subsection{Code overview and input parameters}
\label{overview}
We developed TEPID starting from the prototypical photoionisation code of \cite{nicastro99}, hereafter N99, further adapted by \cite{krongold13}, hereafter K13, to the case of GRB X-ray afterglows. The code solves a set of first-order differential equations following the gas energy (cooling - heating) and ion-fraction (ionisation - recombination) balances as a function of time for the ten astrophysically most abundant elements, i.e. H, He, C, N, O, Ne, Mg, Si, S, Fe. It does it self-consistently and with an in-flight optimised time-resolution, returning, at each instant and at each position in the cloud (see below), the gas ionisation balance, temperature $T$ and transmitted absorption spectrum. 

We follow the ionisation of the cloud in space, from its illuminated face outwards, by implementing a first-order radiative transfer calculation. We divide the gas cloud into a series of optically-thin adjacent shells. The time-evolving calculation is performed shell by shell, starting from the illuminated face of the cloud, from $t=0$ to $t=t_{end}$ and saving all relevant quantities at each time resolution element. The calculation is then repeated at all times for the adjacent shell by using as input, at each time, both the quantities saved at the previous spatial iteration and the geometrically-diluted and opacity-attenuated ionising flux emerging from the previous shell and impinging the shell under consideration (see Eq. \ref{opacity} below).
The calculation is then iterated up to the last shell. The gas temperature and ionisation balance are saved as output of the run, for each time step and for each shell. 
Finally, the ionic population of the gas cloud is fed to a customised version of the PHASE spectral interface \citep{krongold03}, which produces the time-resolved absorption spectra, including both continuum and line absorption, to be compared with observations within the {\it xspec} spectral fitting package.

For each simulation run the input parameters of the code are:
\begin{itemize}
\item $n_e^{tot}$, the total (bound+free) electron density;
\item the Spectral Energy Distribution (SED). TEPID works in the spectral range $10^{-5}$ - 100 keV, i.e. from the infrared, important for Compton cooling (see Eq. \ref{compton} below) up to the hard X-rays to properly account for ionisation and heating;
\item the initial ion abundances $n_{X^i}$ at $t=0$. They can be specified either by providing a value of the ionisation parameter $U$ or through a set of equilibrium ion abundances;
\item $F_{ion} (t)$, the radiative flux impinging on the gas cloud as a function of time, i.e. the lightcurve.
\end{itemize}
The code assumes spherical symmetry and can work in two geometrical settings. In the plane-parallel scenario the geometrical thickness of the shell is negligible with respect to the distance from the luminosity source, i.e. $\Delta r/r <<1$ and, thus, the initial $T, n_{X^i}$ are constant throughout the shell. When such condition is not fulfilled, i.e $\Delta r/r \gtrsim 1$, the input $T, n_{X^i}$ are referred to the illuminated face of the cloud and are then propagated radially outwards throughout the gas column. Elemental abundances are also free parameters: we assume solar metallicity \citep{lodders03} throughout the paper.

We note that at each spatial location (i.e. at each $i$-th shell) all our times $t$ are actually relative times $dt =  t - t_{prop}^i$, where $t_{prop}^i$ is the light travel time from the illuminated face of the cloud to the $i$-th shell. These are actually also the "observer times", i.e., at the relative time $dt$, the observer will see photons that have been emitted by the source at time $t - t^{tot}_{prop}$ (where $t^{tot}_{prop}$ is the light travel time of the entire gas cloud) and that will have already crossed the entire gas column along our line of sight on their way to us.

By the point of view of the transmitted spectrum along the line of sight, all it matters is that at each $i$-th shell the calculation is performed at relative times $dt = t - t^i_{prop}$. The time coordinate in each shell is set by imposing $dt=0$ at $t=0$, i.e. correcting for $t^i_{prop}$. For $dt<0$ (i.e., $t<t^i_{prop}$) the $i$-th shell has not yet been reached by the first photons of the time evolving computation and its gas lies in the status described by the initial conditions (i.e. initial temperature and ion abundances, see \S \ref{initial conditions}). We refer to \cite{kk95} for further discussion on this point.

The new features of TEPID with respect to N99 and K13 are: i) the self-consistent computation of the cooling-heating balance, ii) the in-flight optimised time resolution, iii) the time resolved absorption spectra, iv) the inclusion of collisional ionisation, dielectronic recombination and updated data for photoionisation and radiative recombination (see \S \ref{sub-equations}).

\begin{figure}[!ht]
\centering
\includegraphics[width=0.9\columnwidth]{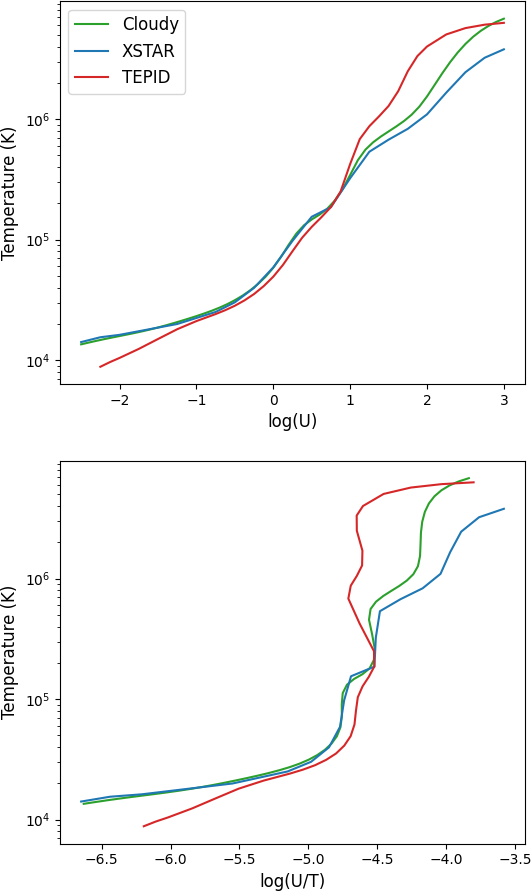}
\caption{Photoionisation equilibrium temperature in an optically thin cloud of gas ionised by the standard AGN SED of Cloudy as a function of the ionisation parameter $U$ (top) and the $U/T$ ratio (bottom). Green, blue and red curves correspond to the values computed with Cloudy, XSTAR and TEPID, respectively.}
\label{UT}
\end{figure}

\subsection{Initial Conditions}
\label{initial conditions}

For a given set of initial conditions the code starts the calculation by guessing a temperature of the gas and computing corresponding "equilibrium" abundances, cooling and heating rates. These are then used to iterate the calculation by varying $T$ until heating and cooling balance each other and, thus, thermal equilibrium is reached. As an example, Fig. \ref{UT} shows the photoionisation equilibrium temperature $T$ as a function of $U$ (top panel) and $U/T$ (i.e. radiation over gas pressure ratio, bottom panel) for an optically thin cloud (log($N_H/$cm$^{-2})=18$) of gas illuminated by the standard AGN SED tabulated in Cloudy\footnote{It consists of a black body component peaking at 10 eV and a double powerlaw at higher energies, with energy index=-1 up to 100 keV and =-2 after.}.
The values computed with TEPID (red line) agree reasonably well with those obtained with Cloudy (green) and XSTAR (blue) over more than three orders of magnitude in $U$ (fully encompassing the range of values typically observed in AGN ionised absorbers, see e.g. \citealp{laha21}), and almost 2 orders of magnitude in $U/T$ (see \S \ref{discuss} for a discussion on the observed differences) \footnote{For this SED, $\xi$ can be derived from $U$ as $log(\xi)=log(U)+1.4$.}.

\subsection{Equations}
\label{sub-equations}
The temporal evolution of the relative abundance of the $i$-th ion of a given element $X$, $n_{X^i}$, can be written as:
\begin{equation}
\begin{split}
& \frac{d n_{X^i}}{dt}= -[F_{X^i} + C_{X^i}n_e + \alpha_{rec}(X^{i-1},T)n_e]n_{X^i} + \\ & (F_{X^{i-1}} + C_{X^{i-1}}n_e)n_{X^{i-1}}
+ \alpha_{rec}(X^i,T)n_e n_{X^{i+1}} + \sum_{k=1}^{i-2}  F^A_{X^k} n_{X^k}
\end{split}
\label{main_eq}
\end{equation}
The first right-hand term is the destruction term due to ionisation from the stage $i$ to $i+1$, both radiative and collisional, with temporal rates $F_{X^i}, C_{X^i}$ respectively, and recombination to the stage $i-1$ with rate $\alpha_{rec}$. The second and third source terms are due to ionisation from the stage $i-1$ to $i$ and recombination from the $i+1$ stage. Finally, the last term accounts for Auger ionisation, which originates from the ionisation of an inner shell electron and leads to multiple ionisation events.
Photoionisation rates are computed by integrating fluxes over the photoelectric cross sections tabulated in Cloudy \citep{Cloudy17} while collisional rates are from \cite{dere07}. $\alpha_{rec}$ is the sum of radiative and dielectronic recombination rates, which are both functions of the gas temperature $T$ and the free electron density $n_e$. For Hydrogen we use the values from \cite{Cloudy96} and we sum over levels from 1 to 20 to get the total recombination rate. For heavier elements total rates to the ground states are taken from: a) \cite{badnell06d,badnell06r} for ions with 0-14 electrons (i.e. fully stripped to Si-like); b) \cite{shull82,m98} for ions with more than 14 electrons (i.e. low charge S and Fe) for radiative and dielectronic recombination, respectively. Auger probabilities are from \cite{kaastra93}.

The gas temperature $T$ is computed by integrating heating and cooling rates of each ion, weighted for the ion abundance $n_{X^i}$ and the element abundance $Z_X$, plus a Compton energy exchange term $\Theta$:
\begin{equation}
\frac{dT}{dt}= \sum_{X,i} ( H(X^i) - \Lambda(X^i) \cdot n_e) \cdot Z_X n_{X^i} + \Theta(F_{ion},T)
\label{temperature}
\end{equation}
$H(X^i)$ consists of two terms:
\begin{equation}
H(X^i)= \int_{\nu_{X^i}}^\infty \frac{F(\nu) \sigma^{X^i}(\nu)}{h \nu} (h\nu-h\nu_{X^i}) d(h\nu) + E_A(X^i) \cdot \sum_{k=1}^{i-2} F^A_{X^k}
\label{heating-term}
\end{equation}
The first term is the heating from photoionised electrons (see e.g. \citealp{netzer13}). $F(\nu)/h \nu$ is the incoming photon flux and it is linked to the ionising flux as $F_{ion}=\int_{10^{-5} keV}^{100 keV} F(\nu) d\nu$. $\sigma^{X^i}(\nu)$ is the ion photoelectric cross section as a function of energy and $(h\nu-h\nu_{X^i})$ is the energy that goes into heating, i.e. the energy of the incident (absorbed) ionising photon minus the binding energy of the electron. The integral is performed from the photon energy threshold $h\nu_{X^i}$ up to the high energy boundary of the SED, i.e. 100 keV. The second term is due to the additional energy $E_A$ carried by the Auger electrons.

Cooling rates for each $X^i$ are taken from \cite{gnat12} and depend both on $T$ and (linearly) on $n_e$. They include line emission, recombination, collisional ionisation and thermal bremsstrahlung.

The net Compton rate $\Theta$ accounts for the energy exchange between the gas and the radiation field via Compton scattering. It can be written as \citep{levich70,Cloudy96}:
\begin{equation}
\Theta=\frac{2 \pi}{m_e c^2} \cdot \frac{n_e}{n_e^{tot}} \Big( \int \sigma_h F(\nu) \ h \nu \ d\nu - 4kT \int \sigma_c F(\nu) d\nu \Big)
\label{compton}
\end{equation}
and includes both heating and cooling terms. In Eq. \ref{compton} $m_e$ is the electron mass and $\sigma_h,\sigma_c$ are the Compton heating and cooling cross sections, respectively.  

The free electron number density $n_e$ is given by the weighted sum over all the elements $X$ and all the ionic levels $i$, each multiplied by the number of ionised electrons:
\begin{equation}
n_e=n_e^{tot} \cdot \frac{\sum_{X} \Big( \sum_{i=2}^{K_X+1} n_{X^i} \cdot (i-1) \Big) Z_X}{\sum_{X} (K_X \cdot Z_X)}
\label{n_e}
\end{equation}
where $K_X$ is the atomic number and $Z_X$ the element abundance.
Finally, the frequency-dependent optical depth within the $m$-th shell, $\tau_m(\nu)$ is given by:
\begin{equation}
\tau_m(\nu)= \delta N_H \cdot \sum_{X} Z_X \Big( \sum_{i=1}^{K_X} n_{X^i} \sigma^{X^i}(\nu) \Big)
\label{tau_eq}
\end{equation}
where $\delta N_H$ is the Hydrogen-equivalent column density of the shell, which is used as a proxy for the total gas column and is obtained by integrating $n_H$ along the line of sight within the shell. The relation between hydrogen and electron number densities depends on the gas metallicity; for solar values (as we use in this paper) $n_H = n_e^{tot}/1.2$. 

Ionisation and heating rates in the $m$-th shell are computed using the spectrum:
\begin{equation}
F'_m=F_m \cdot \frac{1-e^{-\tau_m}}{\tau_m}
\label{eq-shield}
\end{equation}
Where $F_m$ is the incident spectrum at the innermost face of the shell. $F'_m$ accounts for the attenuation of the continuum across the $m$-th shell (see e.g. \citealp{Cloudy17}). 
As discussed in Appendix \ref{app-shield}, using $F'_m$ instead of $F_m$ allows to follow with much higher accuracy the propagation of the radiation through the gas column.

Finally, the incident spectrum on the $(m+1)$-th slab is equal to the transmitted one from the $m$-th one ($F_m \cdot e^{-\tau_m}$) corrected for geometric dilution, i.e. the squared ratio between the radial locations $r_m,r_{m+1}$ of the two slabs:
\begin{equation}
F_{m+1}=F_m \cdot e^{-\tau_m} \cdot \Big( \frac{r_m}{r_{m+1}} \Big)^2
\label{opacity}
\end{equation}

Eqs \ref{main_eq} - \ref{opacity} all depend explicitly on time, and form a coupled system. Therefore, in our code we integrate these equations simultaneously as a function of time. The following subsections show how these equations are solved to obtain the temporal evolution of the gas.

\subsection{Adaptive time binning}
\label{adaptive time binning}

\begin{figure}
\centering
\includegraphics[width=\columnwidth]{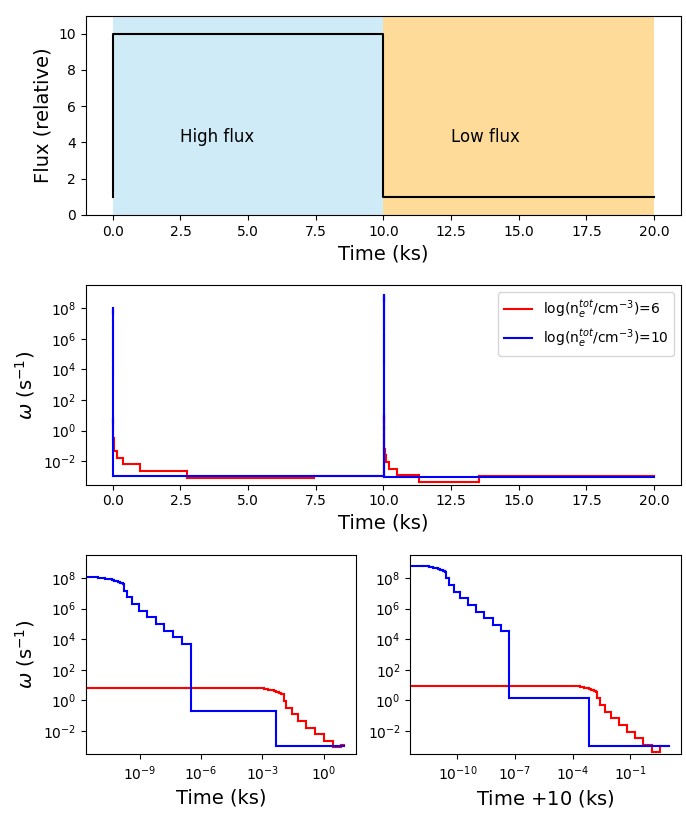}
\caption{Top panel: Step-function lightcurve. Centre: sampling frequency $\omega$. Bottom: zoom-in of $\omega$ around the flux transition phases. See text for details.}
\label{AGN_light+res}
\end{figure}

In order to optimise memory allocation and running time, the temporal sampling frequency $\omega$ is self-evaluated by the code through an adaptive approach. 
In the following example we use an ideal two-phase step-function lightcurve (Fig. \ref{AGN_light+res}, top panel), in which at $t=0$ the gas is in equilibrium with the ionising flux. The flux then increases instantaneously ($d F_{ion}/d t=\infty$) by a factor of 10 and stays at this level for 10 ks (shaded light blue region of top panel of Fig. \ref{AGN_light+res}). At $t=10$ ks the flux abruptly goes back to its initial value and stays at this level for 10 additional ks (shaded orange region) up to $t_{end}=20$ ks. Given the dependence of the equilibration time $t_{eq}$ on $n_e$ (Eq. \ref{teq}), the gas reacts to the luminosity changes at $t=0$ and $10$ ks with a timescale inversely proportional to $n_e$. At high densities the gas adjusts quickly to the abrupt changes in flux and so the time-resolution of the integration has to be high, while at low density the gas takes longer to adjust to the new flux values and lower time-resolutions are sufficient even at the flux transitions phases. This is done automatically through an algorithm, as shown in the middle and bottom (zooms around the flux transition phases) panels of Fig. \ref{AGN_light+res}, that show the frequency $\omega$ of the time-integration sampling for two dramatically different volume electron densities, $log(n_e^{tot}/cm^{-3})=6$ (red curves) and 10 (blue curves).

The algorithm works as follows: 
\begin{itemize}
\item The code scans the entire ionising lightcurve, from $t_{start}$ to $t_{end}$, and divides the time domain into a number of $N_{gross}$ intervals defined by looking for positive or negative flux changes $|\Delta F_{ion} \ge| 10$\%: every time this condition is met, a new time-interval is defined. 
\item For each time interval $n_{gross}^i$ (where $1 \leq n_{gross}^i \leq N_{gross}$), with width $\Delta t^{i}_{gross}$, the equilibration time of each ion is evaluated, using the initial set of equilibrium ion abundances. The longest of these timescales $t_{eq}^{max,i}(n^i_{gross})$ (typically $<< \Delta t^{i}_{gross}$) is used to define, within each $n_{gross}^i$ interval, finer sub-grids each with $N_{fine,i}$ intervals with sizes $\Delta t^j_{fine,i} = Int(\Delta t_{gross}^i / t_{eq}^{max,i}) \times 2^j$, with $j=0, N_{fine,i}$. 
\item The total number of finer intervals from $t_{start}$ to $t_{end}$ is given by $N_{grid} = \sum_{i=1}^{N_{gross}} N_{fine,i}$. This series of interval defines the time-dependent temporal resolution $\Delta t^j_{fine,i}$ of the simulation and, thus, the grid over which the relevant outputs of the computation  (temperature, cooling and heating rates, fractional ion abundances, optical depth and transmitted spectrum) are stored in memory for the next spatial iteration. $\Delta t^j_{fine,i}$ is computed according to the formula in \cite{gallavotti86} to ensure the numerical error of each time step is kept below a certain threshold.
\item The inverse of the temporal resolution defines the time-dependent sampling frequency $\omega_j^i = 1/\Delta t_{fine,i}^j$, i.e. the quantity plotted in Fig. \ref{AGN_light+res} (middle and bottom panels).
\end{itemize}
We verify that the above algorithm provides adequate resolution by re-running the simulations presented in this paper with $\omega_j^i$ increased by a factor 10 and checking that all the results (i.e. temperature, ionic populations, transmitted spectra) are unaffected.

\subsection{General working scheme}
\label{general}
The code structure can be summarised as follows: 
\begin{itemize}
\item After the time optimisation algorithm has defined the sampling frequency $\omega(t)$ (see \S \ref{adaptive time binning}), the time integration starts at the inner, illuminated face of the gas cloud, $r=r_{in}$, and temperature, ionisation and recombination rates within the first slab are updated at each time step and stored in memory.
\item Ion abundances and column densities are printed to an output file.
\item Once $t=t_{end}$ the simulation moves to the second slab. The time integration is repeated by using as incident spectrum that emerging from the previous adjacent slab, corrected, at each temporal resolution step $\Delta t^j_{fine,i}$, for absorption and for geometric dilution according to Eq. \ref{opacity}.
Output files for the ion abundances and column densities are updated with the same temporal resolution as above.
\item The spatial iteration continues until the input Hydrogen-equivalent column density $N_H$ (or outward radius $r_{out}$) is reached.
\end{itemize}
The final outputs of the simulation are $T, n_{X^i}, N_{X^i}, N_H$ as a function of time $t$. We use these quantities to build PHASE table models \citep{krongold03} containing time-resolved UV and X-ray absorption spectra emerging from the cloud. These can be then used in \textit{xspec} to perform time-resolved spectroscopic analysis of the observed data. 

\begin{figure}
\centering
\includegraphics[width=\columnwidth]{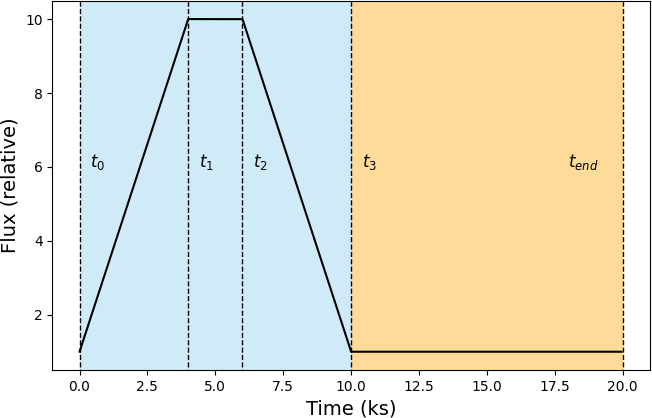}
\caption{AGN lightcurve.}
\label{AGNlightc}
\end{figure}

\begin{figure}
\centering
\includegraphics[width=0.9\columnwidth]{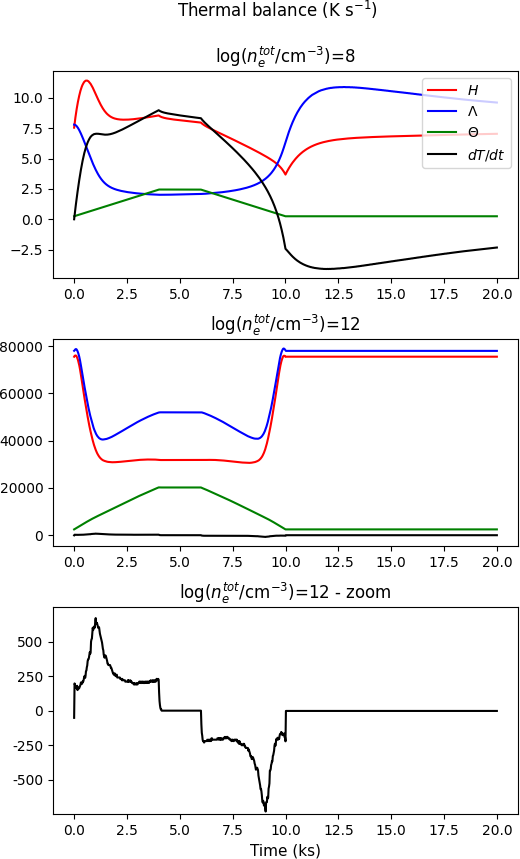}
\caption{Heating $H$, cooling $\Lambda$ and Compton net heating $\Theta$ (red, blue and green lines, respectively) in units of $K\ s^{-1}$ as a function of $t$. Black line represents the algebraic sum of these terms, corresponding to the time derivative of the temperature. Top and centre panels correspond to $\log(n_e^{tot}/cm^{-3})=8,12$, while bottom panel is a zoom-in of the temperature derivative for $\log(n_e^{tot}/cm^{-3})=12$.}
\label{tevol_ft}
\end{figure}

\begin{figure}
\centering
\includegraphics[width=0.8\columnwidth]{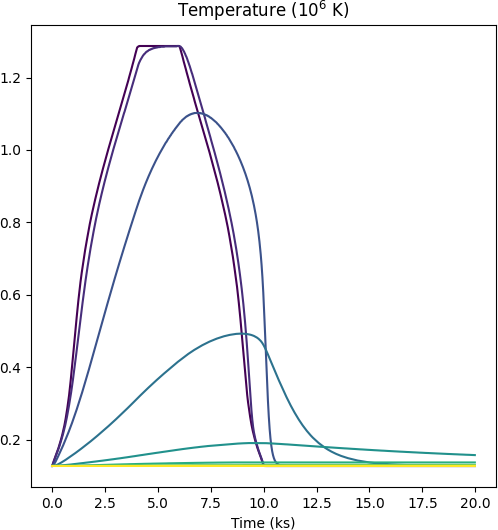}
\caption{$T$ as a function of time for $log(n_e^{tot}/cm^{-3})$ from 4 to 12 (lightest to darkest line).}
\label{tevol_tele}
\end{figure}

\begin{figure*}
\centering
\includegraphics[width=1.8\columnwidth]{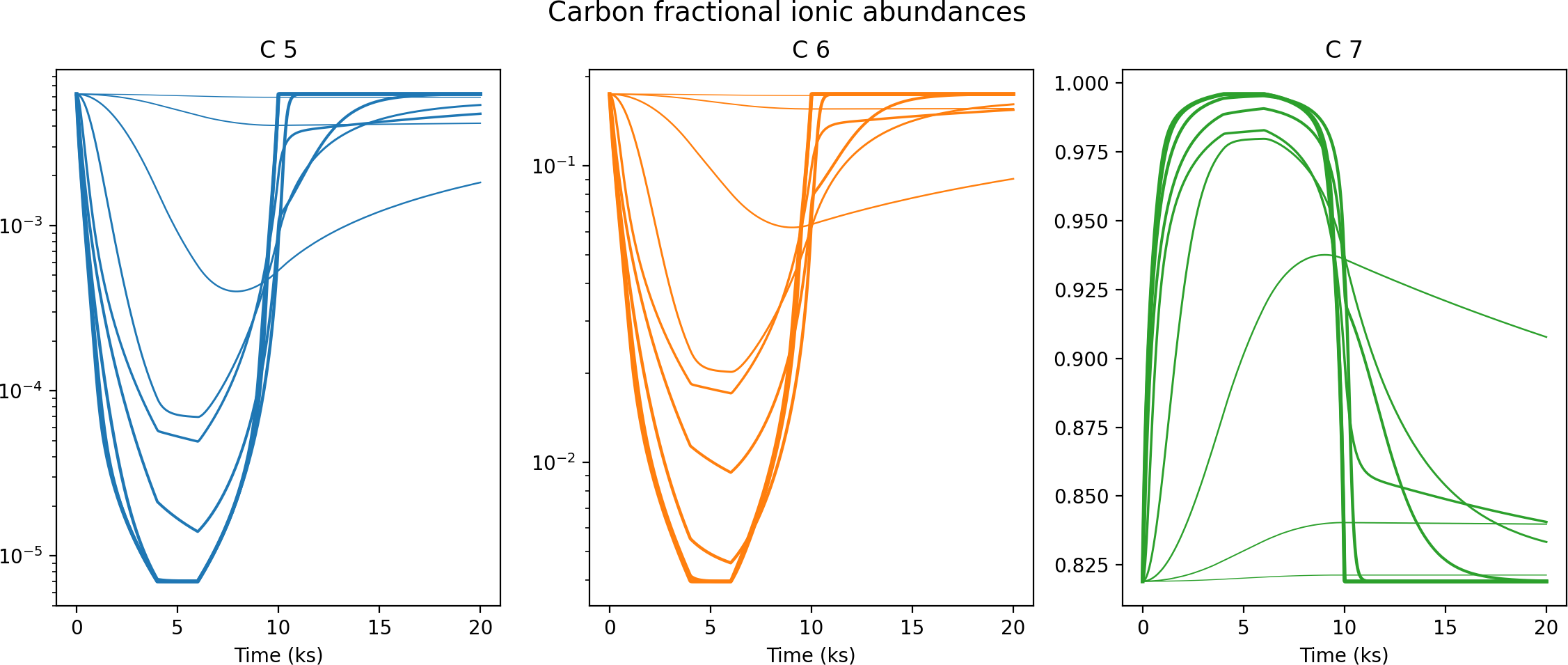}
\includegraphics[width=1.8\columnwidth]{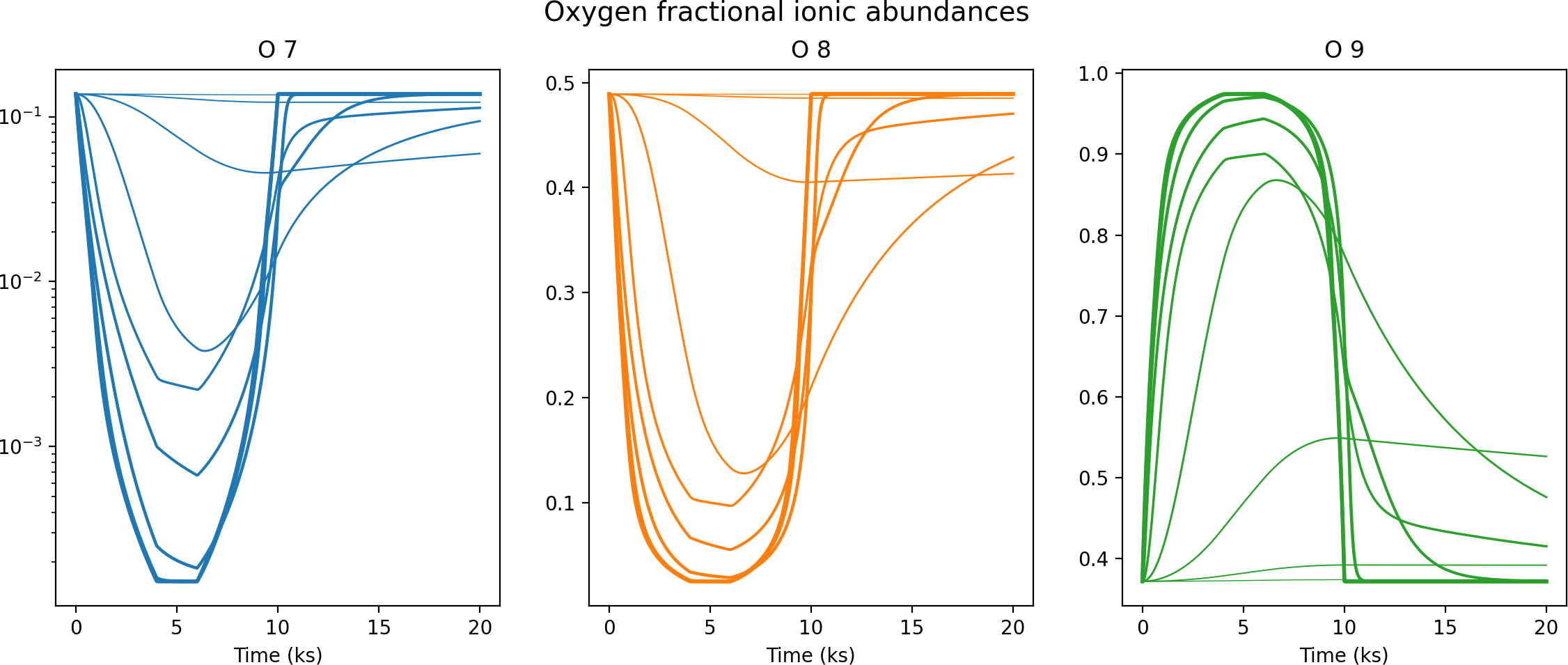}
\includegraphics[width=1.8\columnwidth]{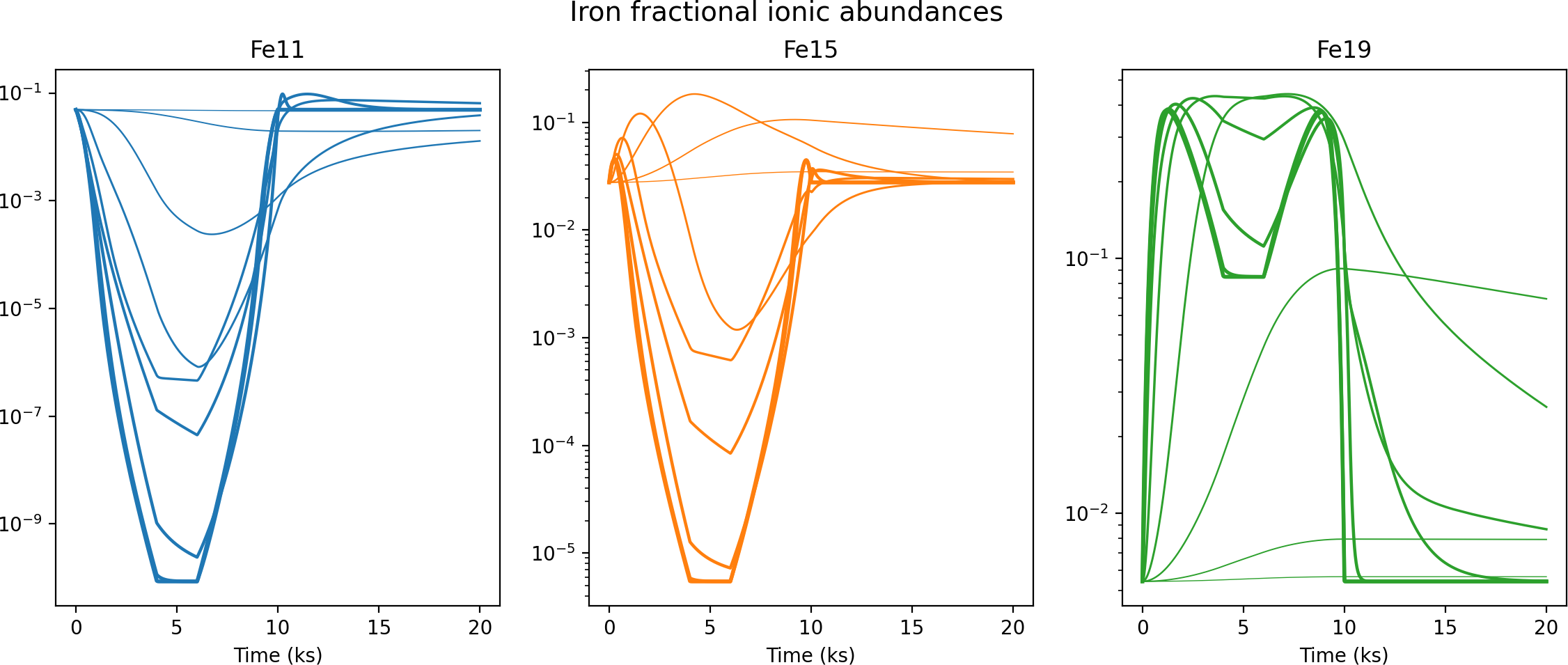}
\caption{Selected ion abundances for Carbon, Oxygen, Iron (top to bottom) as a function of time for $log(n_e^{tot}/cm^{-3})$ from 4 to 12 (thinnest to thickest line). Different panels correspond to different ions, see titles. Note that the y-scale is different for each panel.}
\label{tevol_abund}
\end{figure*}

\subsection{Temperature balance and ion abundances}
\label{agn code}

To illustrate how temperature and ionisation balance evolve under varying ionising flux, here we adopt a plane-parallel setting, where a thin layer of gas ($log(N_H/cm^{-2})=18$) is in photoionisation equilibrium with the radiation field at $t=0$. We assume $log(U)=0.5$ and the standard Cloudy AGN SED described in \S \ref{initial conditions}.
We employ the light curve shown in Fig. \ref{AGNlightc}, i.e. a trapezoidal function with the flux $F_{ion}$ increasing linearly from $F_{min}=1$ (in arbitrary units) at $t=0$ up to $F_{max}=10$ at $t_1$=4 ks, staying at $F_{max}$ from $t_1$ through $t_2$=6 ks and then decreasing again linearly (and with the same slope) to $F_{min}=1$ from $t_2$ through $t_3$=10 ks, and remaining at $F_{min}$ for the next 10 ks (up to $t_{end}$=20 ks).

Figs. \ref{tevol_ft} shows the temporal evolution of $H, \Lambda, \Theta$ for two values of $log(n_e^{tot}/cm^{-3})=8$ (top panel) and 12 (middle and bottom panels), while Fig. \ref{tevol_tele} and Fig. \ref{tevol_abund} show the temporal evolution of the temperature $T$ and of the fractional abundances of the most abundant ions of Carbon, Oxygen and Iron, respectively, for values of electron densities in the range $log(n_e^{tot}/cm^{-3})=4-12$. Generally, for all densities, lower ionisation stages (left and central panels of Fig. \ref{tevol_abund}) are depleted during the high-flux phase and are increased during the low-flux phase, while high ionisation stages (right panels of Fig. \ref{tevol_abund}) show the opposite trend.

In the highest density case, $log(n_e^{tot}/cm^{-3})=12$, the gas is in photoionisation equilibrium and readjusts 'instantaneously' to the incident flux. All quantities follow tightly the lightcurve. In particular, the time derivative of $T$ (i.e. the algebraic sum $H - \Lambda + \Theta$, see bottom panel of Fig. \ref{tevol_ft} for a zoom in) is $>0$ (net heating) during flux increases (from $t_0$ to $t_1$), then is $=0$ during the maximum constant flux phase (from $t_1$ to $t_2$, where the temperature is constant and equal to $10^6$ K, the equilibrium temperature corresponding to $F_{max}$: see Figs. \ref{UT}, \ref{tevol_tele}) and, finally, is negative (net cooling) during the flux decreasing phase (between $t_2$ and $t_3$).
Accordingly, the temporal evolution of the ion abundances (Fig. \ref{tevol_abund}) closely follows the lightcurve of the ionising source by quickly adjusting to the instantaneous equilibrium values.

For decreasing $n_e^{tot}$ the evolution of the gas physical properties departs from equilibrium. Ion-by-ion photoionisation rates do not explicitly depend on $n_e^{tot}$; however, for a given initial $log(U)$ the incident flux $F_{ion}$, which is proportional to photoionisation rates, scales linearly with $r^2 n_H \propto r^2 n_e^{tot}$. So, at a given distance, lower values of $n_e^{tot}$ correspond to lower photoionisation rates. Radiative and dielectric recombination rates, instead, explicitly depend on $n_e$ and also on $T$ via a polynomial fitting formula (see refs. in \S \ref{initial conditions}).

This leads to two time-dependent effects for decreasing $n_e^{tot}$. First, both recombination and photoionisation rates are slower and thus the variations of $T$ and $n_{X^i}$ in response to the flux changes are not only compressed, but can be delayed with respect to the changes. This, in turn, affects further recombination rates, which are slower not only because of the lower density but also due to the compressed dynamics of temperature variations. The delayed response of the gas is clearly visible in Fig. \ref{tevol_tele} and \ref{tevol_abund}, where the maximum $T$ and fractional abundance of fully stripped ions are not simultaneous to the peak of the lightcurve (i.e. between $t_1$ and $t_2$) but are increasingly shifted toward the end of the high flux phase ($t=10 ks$) as density decreases, and the gas stays over-heated and over-ionised at $t>10 ks$. This is also evident in the top panel of Fig. \ref{tevol_ft} ($log(n_e^{tot}/cm^{-3})=8$), where  $dT/dt$ assumes negative values at $t>10 ks$ (i.e. net cooling, since the gas temperature is higher than the equilibrium one). As we will show below, these effects lead to unique spectral signatures in time-resolved spectra of variable ionising sources that, when properly modelled, allow the characterisation of the gas electron density $n_e^{tot}$.

\begin{figure*}
\centering
\begin{tabular}{cc}
\includegraphics[width=8.5cm]{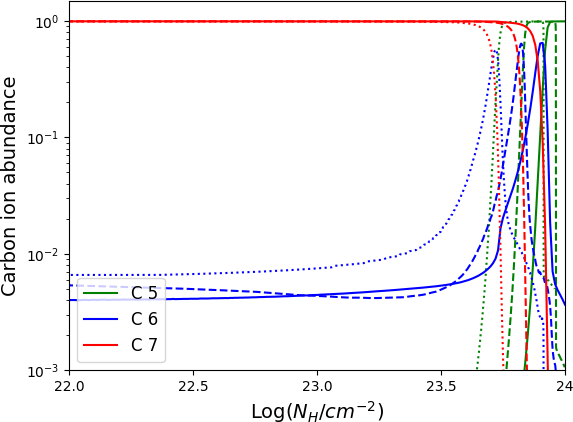} & \includegraphics[width=8.5cm]{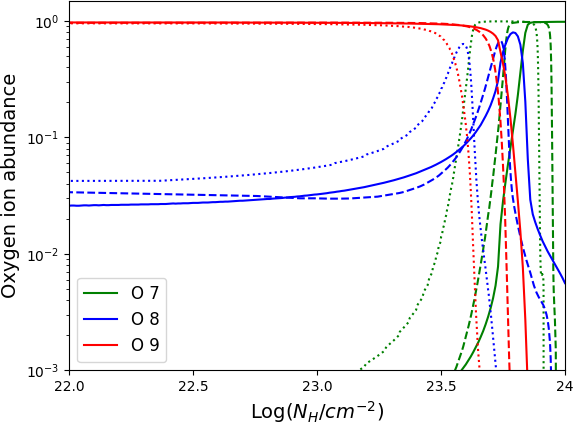} \\
\includegraphics[width=8.5cm]{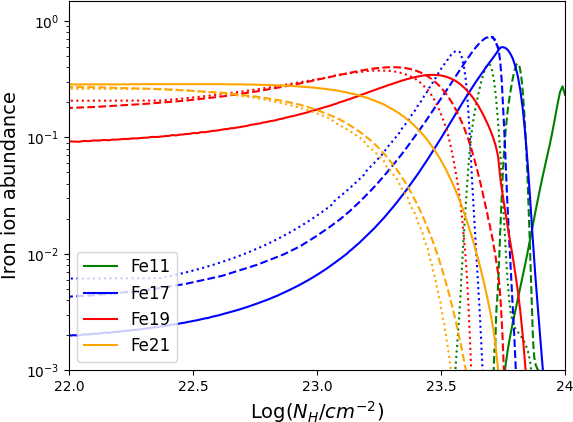} & \includegraphics[width=8.5cm]{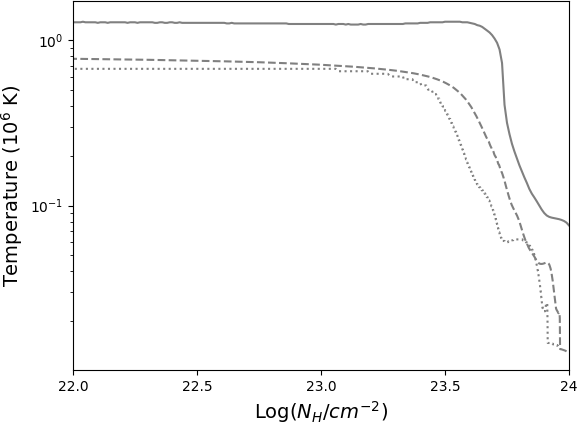} \\
\end{tabular}
\caption{Selected ion abundances (from top to bottom and left to right, Carbon, Oxygen, Iron; see legends) and temperature (bottom right) as a function of $N_H$ for a gas in photoionisation equilibrium, computed with TEPID (solid lines), Cloudy and XSTAR (dashed and dotted lines, respectively).}
\label{Cloudy-comp}
\end{figure*}

\begin{figure*}
\centering
\includegraphics[width=10cm]{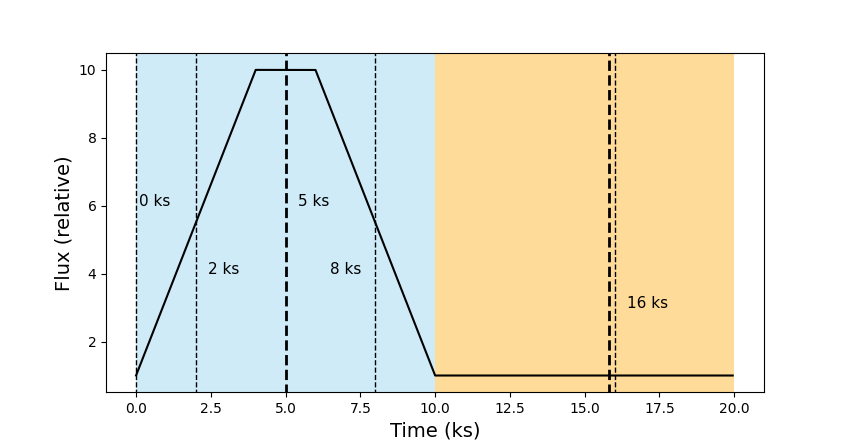}
\includegraphics[width=1.5\columnwidth]{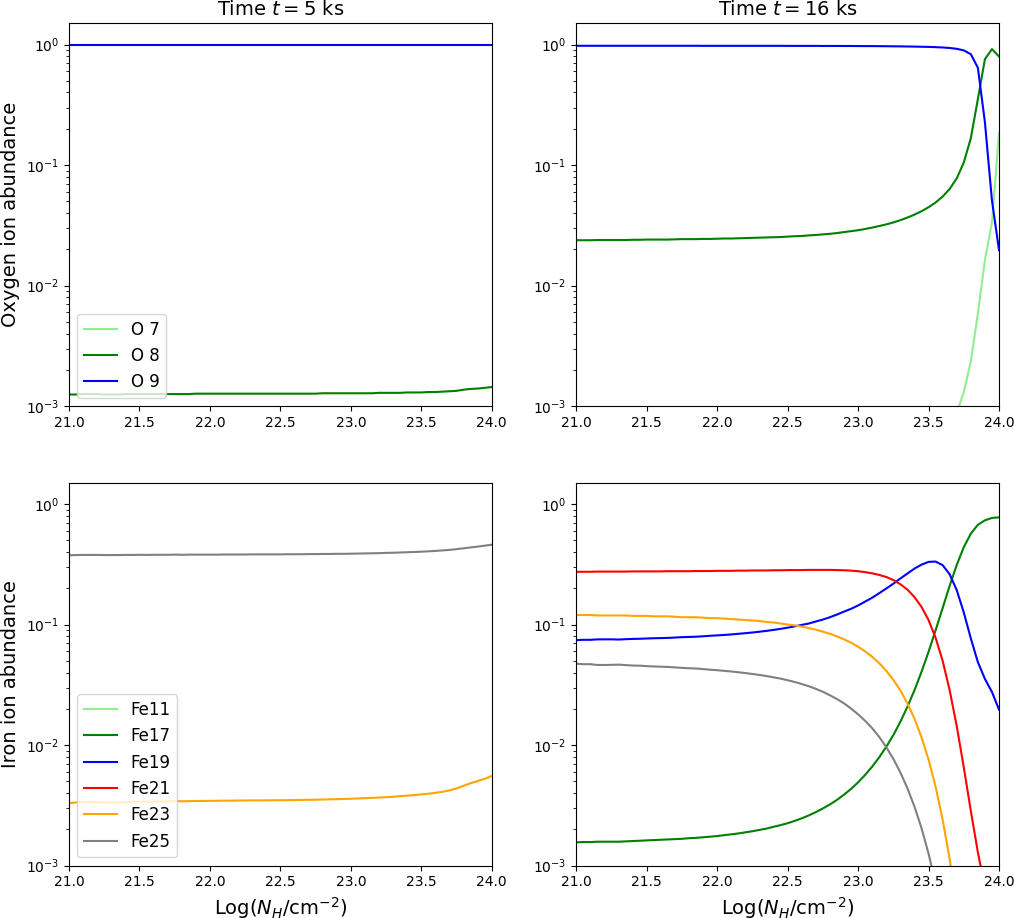}
\caption{Top: AGN lightcurve. Thick lines mark the times for which ion abundances are shown (see below). Thin lines show the times corresponding to the absorption spectra in Fig. \ref{AGN_spectra_t-n}. Bottom, first and second row: selected ion abundances (see legend) respectively for Oxygen and Iron as a function of $N_H$. Left and right panels correspond to $t=5, 16 ks$.  $log(n_e^{tot}/cm^{-3})=10$ in all cases.}
\label{tevol_teleNH}
\end{figure*}

\begin{figure*}
\centering
\includegraphics[width=13cm]{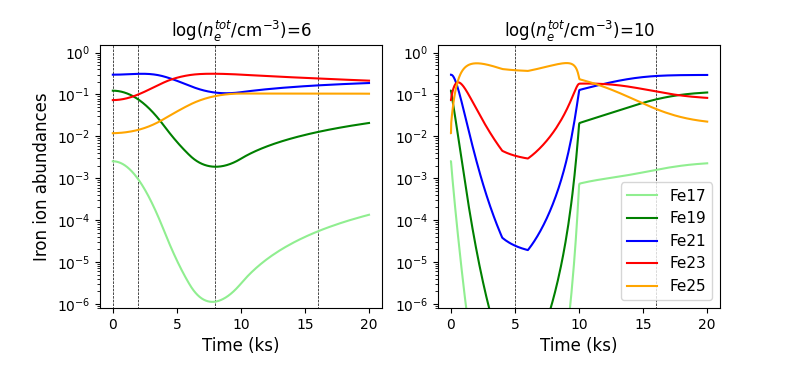}
\includegraphics[width=13cm]{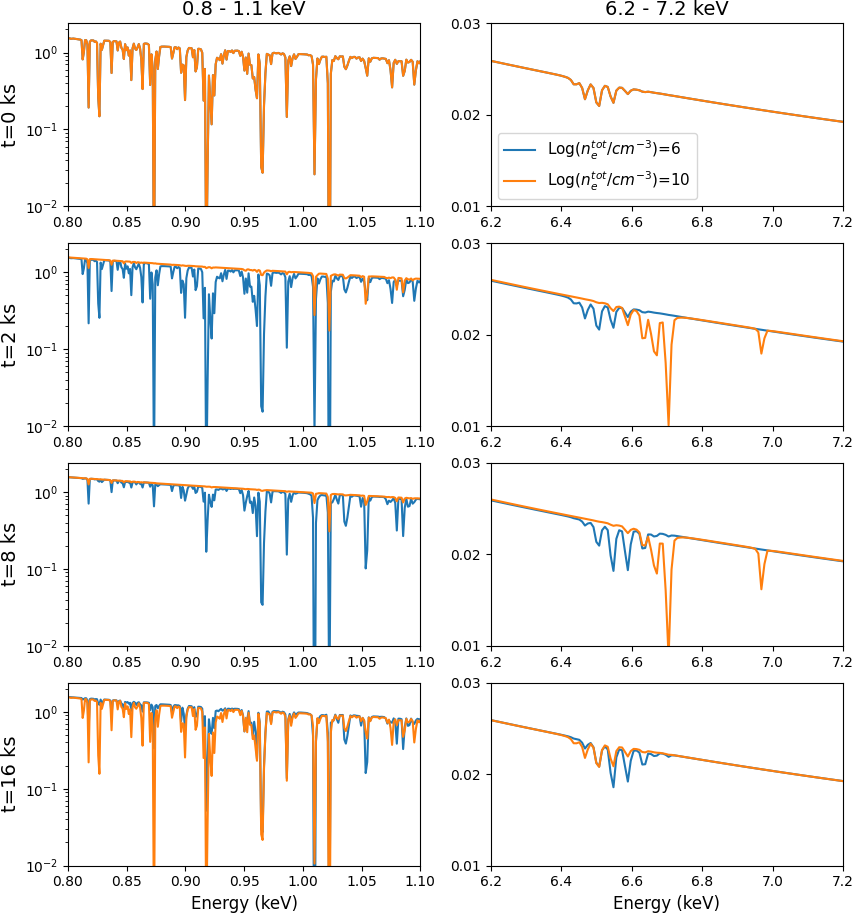}
\caption{Top: selected Iron ion abundances (see legend) as a function of $t$ for $log(n_e^{tot}/cm^{-3})=6,10$ (left and right panel). Dashed lines mark the times corresponding to the spectra shown below. Bottom: Absorption spectra for $\log(N_H/cm^{-2})=22$ and t=0,2,8,16 ks (top to bottom, corresponding to the times marked above and in Fig. \ref{tevol_teleNH}) and $\log(n_e^{tot}/cm^{-3})=6,10$ (blue and orange lines). Left and right columns show respectively the Fe L and Fe K bands (i.e. 0.8-1.1 and 6.2-7.2 keV).}
\label{AGN_spectra_t-n}
\end{figure*}

\section{Time-Resolved Spectra of Ionised AGN Absorbers}
\label{agn-out}

In this section we show an application of TEPID to the case of highly ionised absorbers in AGNs. We set the AGN SED of \S \ref{agn code} for the ionising source and an initial equilibrium ionisation parameter of the gas to $log(U)=1.5$ (a factor 10 higher than in \S \ref{agn code}), typical of both the high-ionisation components typically observed in Warm Absorbers (see e.g. \citealp{krongold03,krongold09}) and the low-ionisation counterparts (e.g. \citealp{krongold21}) of Ultra-Fast Outflows (UFOs, e.g. \citealp{tcr10,lne21}). 
Simulations are run for a range of densities $log(n_e^{tot}/cm^{-3})=4-12$ and up to an equivalent hydrogen column density $log(N_H/cm^{-2})=24$. As a consistency check of our radiative transfer (see Eqs. \ref{eq-shield}, \ref{opacity}), we first run TEPID at equilibrium (by inputting a constant lightcurve) and compare the resulting ion abundances and temperature with Cloudy. Then, we use the same lightcurve of Fig. \ref{AGNlightc} to illustrate the time-dependent behaviour.

\subsection{Photoionisation equilibrium}
\label{phot-eq}
Fig. \ref{Cloudy-comp} shows the ion abundances for Carbon, Oxygen, Iron and $T$ as a function of $N_H$ for a gas in photoionisation equilibrium with $log(U)=1.5$. Solid, dashed and dotted lines correspond to TEPID, Cloudy and XSTAR, respectively. At low column densities the gas is quite transparent: its ionisation and temperature are that of an optically thin gas up to $log(N_H/cm^{-2}) \approx 22.5$. The agreement between the ionic populations of TEPID and Cloudy is within a factor 2 for the main ions (i.e. those with abundance $> 10^{-2}$). At higher $N_H$, the TEPID radiative transfer correctly accounts for the increasing optically thickness of the gas and the (decreasing) gas ionisation and temperature are in good agreement with the Cloudy and XSTAR ones.

\subsection{Time-dependent Optical Depth}
Fig. \ref{tevol_teleNH}, second and third row, shows the ionic fractions of O, Fe as a function of $N_H$ for $log(n_e^{tot}/cm^{-3})=10$ and at two different times during the simulations: $t_1$=5 ks and $t_2$=16 ks (see caption). With an initial $logU=1.5$, H, He and light metals are fully ionised at virtually all times during the simulations and the gas becomes optically-thick only at $log(N_H/cm^{-2}) \gs 23$. Consequently, the gas column is virtually isothermal at all times during the simulation (i.e. $T(t_1)=3\cdot 10^6$ and $T(t_2)=10^6$) and has constant light metal abundances (i.e. $f_{C VII} \simeq f_{O IX} \simeq 1$) up to $log(N_H/cm^{-2}) \simeq 23$ (middle panel of Fig. \ref{tevol_teleNH}).

At $log(N_H/cm^{-2}) \gs 22.5 - 23$ (depending on the considered element), the gas become more and more optically thick and both its temperature and ionisation degree decrease steadily. However, the smoothness of this decrease depends both on $n_e^{tot}$ and on the ionising flux state. At the chosen $log(n_e^{tot}/cm^{-3})=10$ and during the minimum flux state ($t=16$ ks), the abundance of the fully-stripped oxygen ion starts decreasing slowly at $log(N_H/cm^{-2}) \gs 23.8$ (right middle panel of Fig. \ref{tevol_teleNH}), while the He-, Be- and C-like fractional abundances of Fe start decreasing quickly already at $log(N_H/cm^{-2}) \gs 22.5$ to the advantage of the Ne- and O-like ions (right bottom panel of Fig. \ref{tevol_teleNH}).

\begin{figure*}[!ht]
\includegraphics[width=2\columnwidth]{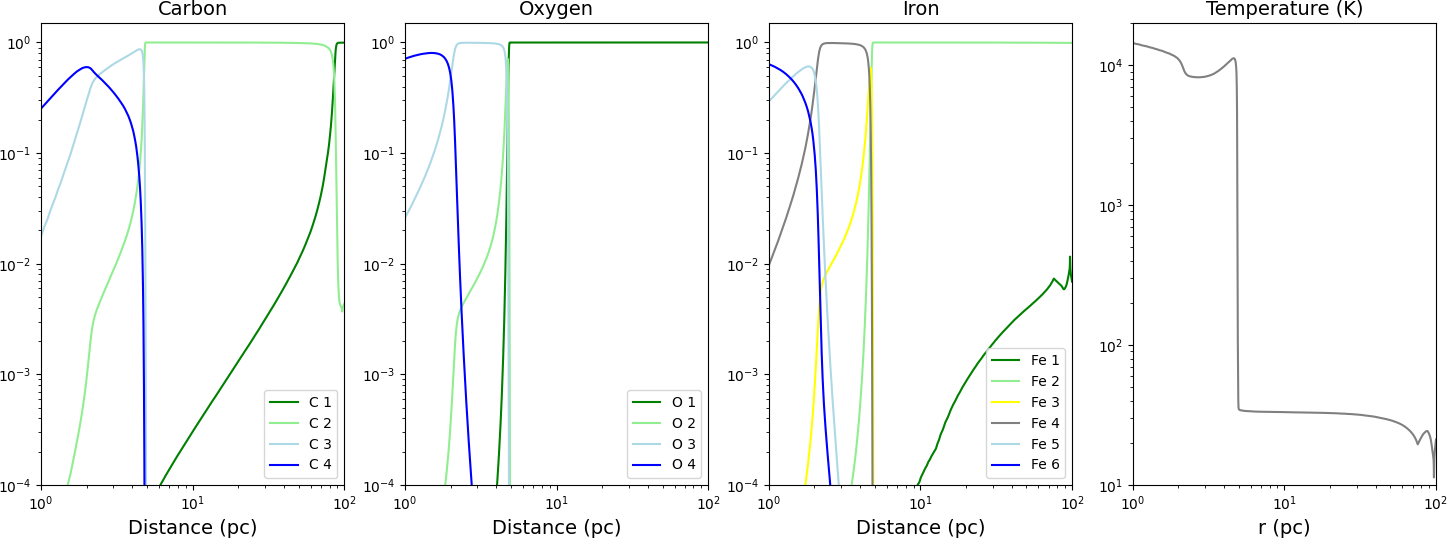}
\caption{Pre-Burst ionisation structure of the surroundings of a Wolf-Rayet star at the centre of a star-forming region with radius 100 pc and constant density $n_e^{tot}=10^2 cm^{-3}$. From left to right, the panels show C, O, Fe ion abundances and Temperature as a function of the distance (in pc) from the star.}
\label{preion-initabunds}
\end{figure*}

\begin{figure}[!ht]
\centering
\includegraphics[width=\columnwidth]{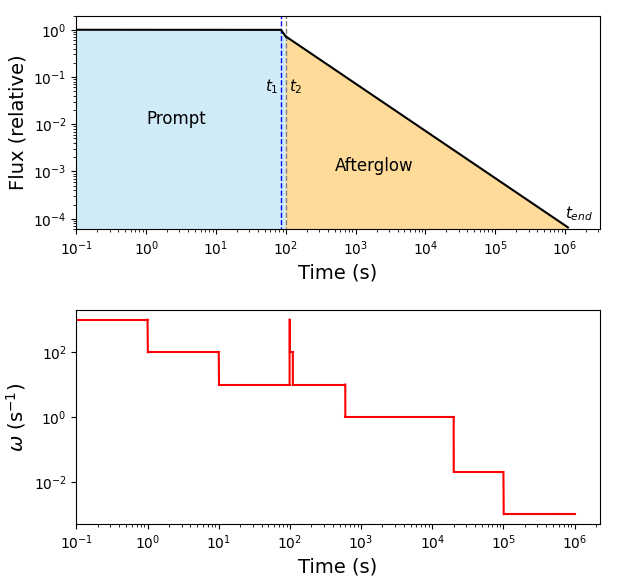}
\caption{Top: GRB lightcurve. $t_1, t_2, t_{end}$ correspond respectively to the end of the initial constant luminosity phase, the start of the afterglow and the end time of the simulation. Bottom: resolution $\omega$ as a function of time.}
\label{GRB_light+res}
\end{figure}

\subsection{Time-Resolved Spectra of AGN Outflows}
Fig. \ref{AGN_spectra_t-n} shows absorption spectra extracted from the above simulation for $log(N_H/cm^{-2})=22$ and $log(n_e^{tot}/cm^{-3})=6,10$ (blue and orange curves, respectively), at different times $t=0, 2, 8$ and 16 ks.
For reference we report in the top panels the Fe ion abundances as a function of time. Left and right panels show, respectively, the spectral regions containing the Fe L (0.8-1.1 keV) and Fe K (6.2-7.2 keV) transitions. At $t=0$ (top panels) the gas is in photoionisation equilibrium, therefore the spectra are independent on $n_e^{tot}$ (orange superimposed to blue). 

At high density ($log(n_e^{tot}/cm^{-3})=10$, orange curves) the gas quickly responds to flux variations, though is never (except at $t=0$) in precise photoionisation equilibrium with the ionising flux; therefore, the $t=2 $ks and $t=8$ ks spectra, corresponding to equal values of $F_{ion}$ symmetrically preceding and following the maximum, respectively, are similar, but not identical. Analogously, the $t=16$ ks spectrum (bottom panels) is similar (though not identical) to the initial equilibrium spectrum, because both are taken at the minimum ionising flux level.

Conversely, when $log(n_e^{tot}/cm^{-3})=6$ (blue curves) the gas is out of equilibrium at all times $t>0$ and the temporal evolution of its ionisation balance and temperature is both compressed and delayed, as discussed in \S \ref{agn code} above. This is clearly reflected in the spectra at $t=2, 8$ and 16 ks. Little or no spectral variation is observed in both spectral bands between $t=0$ and $t=2$ ks, despite the increase of the ionising flux. Between $t=2$ and $t=10$ ks the gas ionisation degree increases steadily despite a first constant ($t=2-4$ ks) and then steadily decreasing ($t=4-10$ ks) ionising flux. Correspondingly the $t=8$ spectrum is over-ionised with respect to its ionising flux and shows lower opacity in its low-ionisation Fe transitions than the spectrum at $t=2$ ks. The gas ionisation keeps increasing even at $t>10$ ks, while the ionising flux stays constant and equal to the minimum flux in the simulation. Consequently, the low density spectrum at $t=16$ ks is the least opaque of the four spectra of Fig. \ref{AGN_spectra_t-n} in its Fe-L energy band.
The gas will eventually get back to equilibrium only after $\approx$ 300 ks the flux has returned to its minim value.

Following the evolution in time of the ionisation degree of AGN outflows with X-ray telescopes offers then a unique opportunity to constrain, via time-resolved spectroscopy, the electron density of the gas and thus its distance from the ionising source. This, in turn, allows one to estimate the mass load of the outflow and its energetic. 

\begin{figure}[!ht]
\centering
\includegraphics[width=\columnwidth]{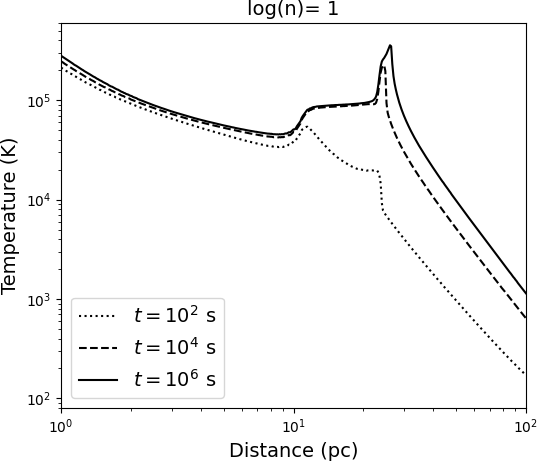}
\caption{Temperature for $\log(n_e^{tot}/cm^{-3})=1$ as a function of the radius $r$ (i.e. the distance from the GRB) for $t=10^2,10^4,10^6$ s (dotted, dashed and solid lines, see legend).}
\label{GRB_tele}
\end{figure}

\begin{figure*}
\centering
\includegraphics[width=1.8\columnwidth]{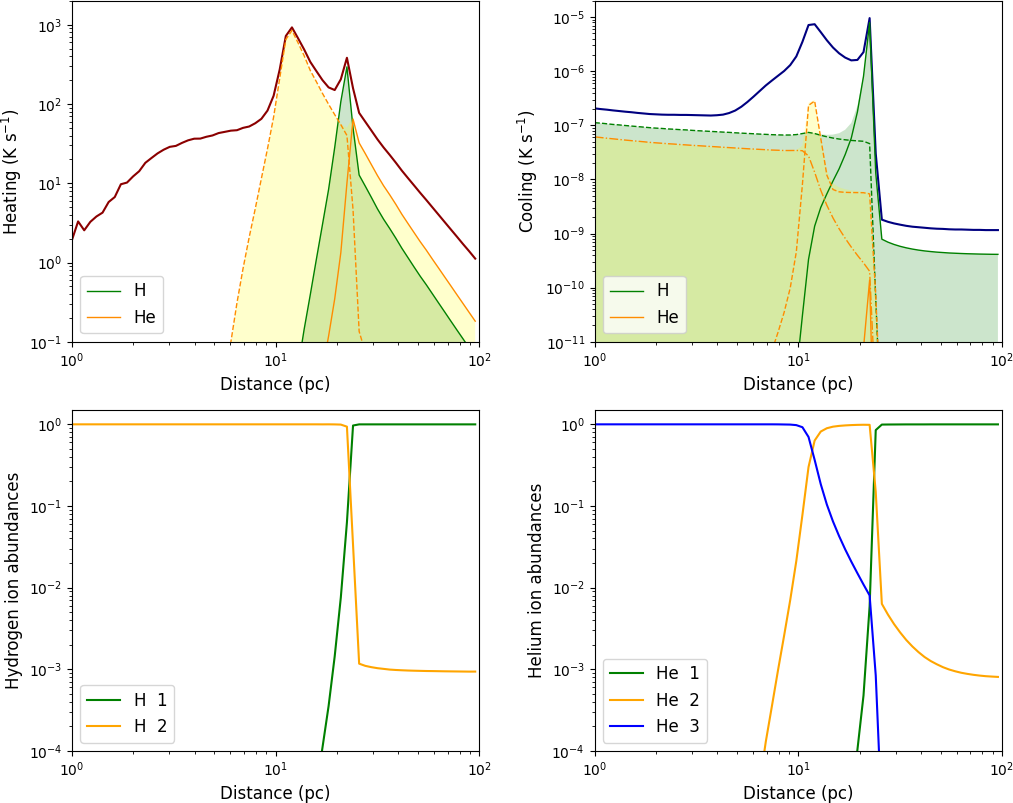}
\caption{Top panels: heating (left) and cooling (right) rates as a function of the radius $r$. Red line in the left panel and blue line in the right panel correspond to the total rates, while green and yellow shaded areas correspond to the total Hydrogen and Helium rates, respectively. Green(yellow) solid and dashed lines correspond to the rates of H(He) I, II, respectively, while yellow dot-dashed line to He III. Bottom: Hydrogen and Helium (left and right, respectively) ion abundances (colour coding, see legend) as a function of $r$. In all cases $\log(n_e^{tot}/cm^{-3})=1,t=10^2$ s.}
\label{recomb_fronts}
\end{figure*}

\begin{figure*}
\centering
\begin{tabular}{ccc}
\includegraphics[width=0.6\columnwidth]{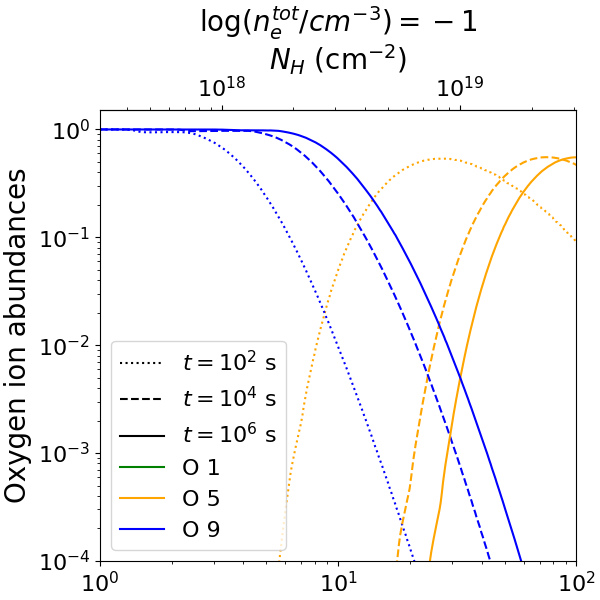} &
\includegraphics[width=0.6\columnwidth]{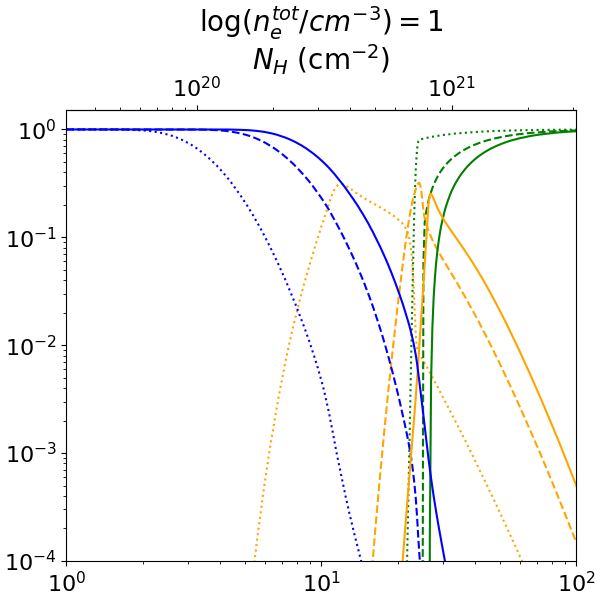} &
\includegraphics[width=0.6\columnwidth]{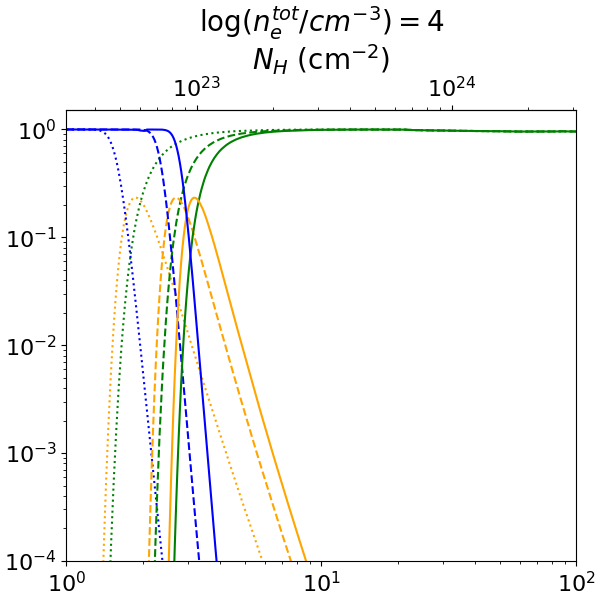} \\
\includegraphics[width=0.6\columnwidth]{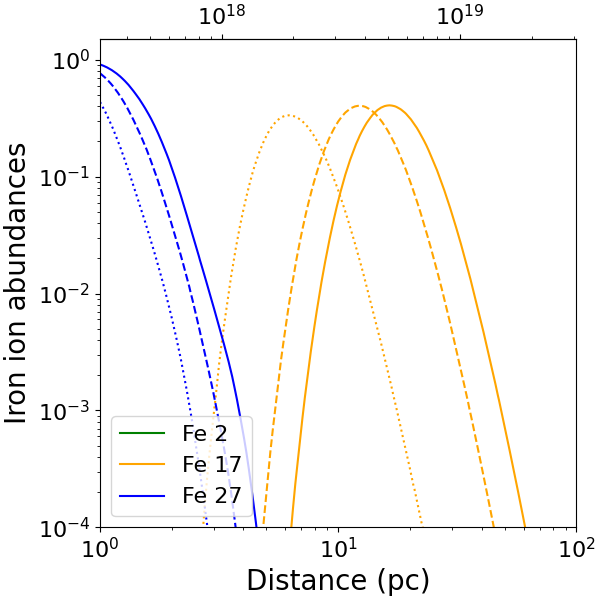} &
\includegraphics[width=0.6\columnwidth]{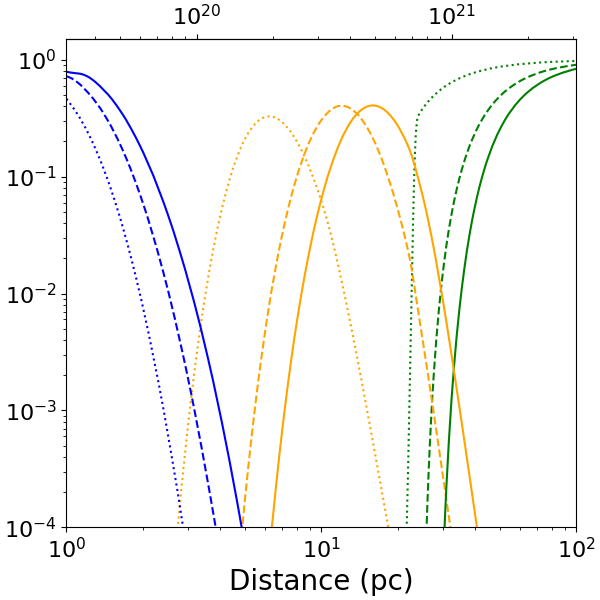} &
\includegraphics[width=0.6\columnwidth]{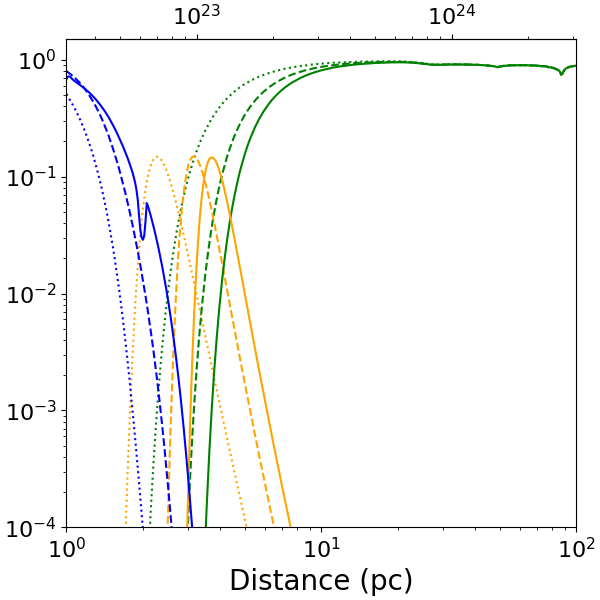} \\
\end{tabular}
\caption{Abundances for some Oxygen (top) and Iron (bottom) ions (colour coding, see legend) as a function of the radius $r$. Dotted, dashed and solid lines identify increasing times from the onset of the GRB, i.e. t=$10^2,10^4,10^6$ s, respectively, while left, central and right panels correspond to increasing ISM density, i.e. $\log(n_e^{tot}/cm^{-3})=-1,1,4$.}
\label{GRB_movie}
\end{figure*}

\section{The ionisation of the surroundings of GRB afterglows}
\label{grb-out}
Here we use TEPID to reproduce the highly structured gaseous ionised environment of long GRBs, thought to originate from the death of massive stars, embedded in a dense star-forming region \citep{perna07,heintz18,fryer21}. Our aim is to show how proper time-evolving photoionisation modelling of GRB X-ray afterglow spectra can efficiently constrain the physical and chemical properties of the interstellar medium of galaxies hosting a GRB explosion. 
\subsection{Physical setting of the GRB surroundings}
\label{phys-set}
For simplicity we assume that the interstellar matter (ISM) surrounding the GRB has constant number density $n_e^{tot}$, i.e., independent on the radial distance $r$ from the GRB. The inner boundary is set to $r_{in} = 1 pc$ to encompass the forward shock, from where much of the X-ray radiation is irradiated. We set an outer radius $r_{out}=100 pc$, i.e. an upper envelope for the size of a typical OB stellar association \citep{stahlerpalla04}, where the massive progenitor star is expected to assemble. Since the density is constant, the total gas Hydrogen-equivalent column density is simply given by $N_H=(r_{out}-r_{in}) \cdot n_H$. Unless otherwise specified, we focus on the density range $\log(n_e^{tot}/cm^{-3})=1-4$.
We also neglect contributions from the outflows associated with the prompt phase or the progenitor star. This is justified since at a distance $r_{in}$ the density of a typical Wolf-Rayet wind is already $\sim 1$ order of magnitude lower than the typical intragalactic densities ($n\simeq$ 1 cm$^{-3}$) and, thus, much lower than the range of densities we explore in our simulations \citep{2022MNRAS.515.2591C, 2006MNRAS.367..186E}.

Finally, for the pre-burst physical conditions of the gas permeating the star-forming region we assume those of a medium photoionised by a bright and hot Wolf-Rayet star, i.e. a blackbody with temperature of $10^5$ Kelvin and a bolometric luminosity of $10^{5.8} L_{\odot}$ (\citealp{perna18}, but see \S \ref{discuss} for relaxations of this hypothesis). Figs. \ref{preion-initabunds} shows the pre-burst most abundant ions of C, O, Fe and the temperature as a function of the radial distance.

\subsection{The ionising source}
\label{ion_sou}
We parameterise the source ionising flux $F_{ion}(t)$, i.e. the GRB spectrum and its temporal evolution, as in K13, based on the analysis of the broad sample of \citet{margutti13} consisting of more than 650 GRBs observed with {\it Swift}.
This has different X-ray spectral shape and lightcurve in the prompt ($t\le 100$ s) and afterglow ($t> 100$ s) phases. 
The X-ray prompt phase is described as a powerlaw spectrum with energy spectral index $\alpha_{xp}=0.0$ (i.e., a flat spectrum), irradiating with constant luminosity up to $t_1=$85 seconds and decaying as $t^{-2}$ between $t_1$ and $t_2=100$ s. At $t_2$ the GRB transits to the afterglow phase: the spectrum becomes steeper, $\alpha_{xag}=1.0$, and the light curve flattens as $t^{-1}$.
Fig. \ref{GRB_light+res}, top panel, shows the lightcurve from $t=0$ up to $t_{end}=10^6$, a typical end time of X-ray afterglow observations. The total irradiated isotropic energies in the range 0.3-10 keV during the prompt and afterglow (up to 10$^6$ s) phases are $1.4 \cdot 10^{51} erg$ and $5.4 \cdot 10^{51} erg$, respectively.
Given the shape of the lightcurve and the extremely high initial flux, the internal time-step optimisation algorithm chooses a finer resolution at earlier times which then gradually decreases with time. We plot the time sampling frequency $\omega$ in the bottom panel of Fig. \ref{GRB_light+res}. It can be seen that $\omega$ decreases from $10^3 s^{-1}$ at $t=0$ to $10^{-3} s^{-1}$ at $t=t_{end}=10^6$ s. The only exception is between 85 and 100 s, where $\omega$ increases to better follow the temporal decay transition.

\subsection{Temperature and ion abundances as a function of time and distance}
\label{GRBout}

\begin{figure*}
\centering
\includegraphics[width=1.6\columnwidth]{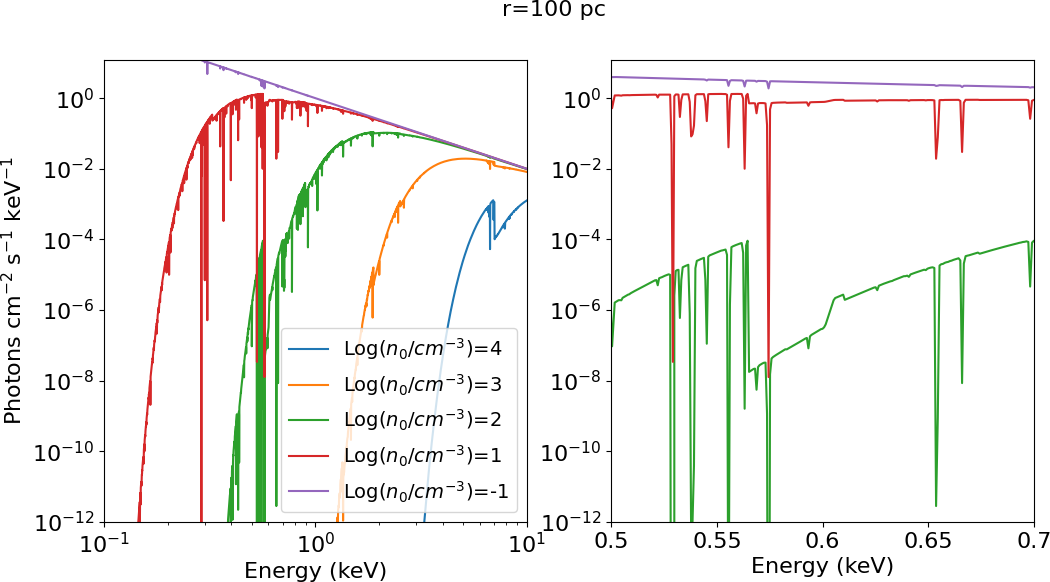}
\caption{Left: absorption spectra for a distance $r=100 pc$ from the GRB and increasing $n_e^{tot}$ (and, thus, increasing $N_H$; colour coding, see legend). Right: a zoom-in on the 0.5-0.7 keV range, showing several oxygen absorption lines from neutral to hydrogen-like states.}
\label{theor_spectra_nh}
\end{figure*}

\begin{figure*}
\centering
\includegraphics[width=1.6\columnwidth]{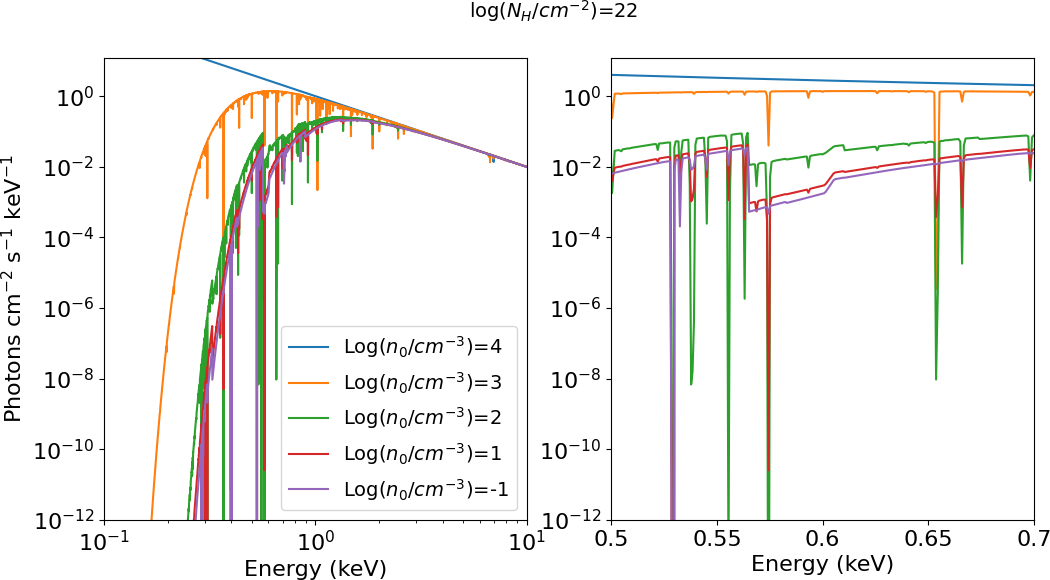}
\caption{Left: absorption spectra for a column density $N_H=10^{22} cm^{-2}$ and increasing $n_e^{tot}$ (and, thus, decreasing distance from the GRB; colour coding, see legend). Right: a zoom-in on the 0.5-0.7 keV range.}
\label{theor_spectra_n}
\end{figure*}

Fig. \ref{GRB_tele} shows the post-explosion temperature of the surrounding ISM as a function of the distance $r$ from the GRB for a density $log(n_e^{tot}/cm^{-3})=1$ and three increasing times (dotted, dashed and solid lines, respectively). 
When the burst explodes ($t=0$) it quickly and fully ionises its immediate surroundings (photoionisation largely dominates over recombination, due to the low density and the 18 orders of magnitude flux increase) and the temperature of the ISM reaches very high values ($\sim 2\times 10^5$ K at $t=100$ s, in this example). The extent of the ionised region expands with time (dotted to solid lines) driven by the ionising flux, which however gets geometrically-diluted and increasingly absorbed for increasing distance. As an example, the flux of ionising photons at $r=20 pc$ is lower than that at $r=10 pc$ by a factor $(20/10)^2=4$, therefore it takes 4 times longer to collect the same number of ionising photons (and, so, to reach the same degree of ionisation within the medium), provided that $F_{ion}$ stays constant, i.e. during the prompt phase. During the afterglow the expansion of the ionised region is even slower, since $F_{ion}$ decrease with time as $\propto 1/t$ (see Fig. \ref{GRB_light+res}). 

The temperature smoothly decreases with distance up to critical radii that depend on the time after the explosion and the density of the ISM (and so its cumulative opacity to the incoming radiation). At these radii the temperature profile displays sudden jumps or discontinuities to then continue decreasing smoothly and eventually approaching the temperature of the pre-burst ISM at large distances. These time-dependent, critical distances are set by the recombination fronts of abundant elements in high-ionisation states (particularly H and He, but also, at a lower extent, lower ionisation species of heavier elements).
These recombination fronts can be clearly seen in Fig. \ref{recomb_fronts}, where the contributions of H, He ions to the heating and cooling rates (top left and right panels) are displayed together with their fractional abundance (bottom left and right panels, respectively), as a function of radial distance from the burst for $t=100$ s and $log(n_e^{tot}/cm^{-3})=1$. The first recombination front is found at a distance of about 11 pc and is due to recombination of HeIII in HeII (bottom right panel), while the second front is located at $\sim 20$ pc and is due to both the recombination of HeII in HeI and, most importantly for cooling, the recombination of H, which becomes virtually all neutral at $r\gs 20$ pc (bottom left panel). Correspondingly, heating and cooling rates show peaks and edges at the recombination fronts. The temporal evolution of such fronts, which move outwards as the medium gets progressively ionised with time, lead to the peaks in the heating and cooling and, thus, in the temperature shown in Fig. \ref{GRB_tele} between $\approx 10$ and 30 parsec.

Finally, Fig. \ref{GRB_movie} shows the abundances of the most abundant ions of Oxygen (top panels) and Iron (bottom panels), as a function of the distance $r$ from the GRB (bottom X-axis) and the line of sight Hydrogen-equivalent column density $N_H$ (top X-axis) for three increasing times (dotted, dashed and solid curves). 
Left to right panels correspond to different values of the ISM density, $log(n_e^{tot}/cm^{-3})=-1,1,4$, respectively. We include  $log(n_e^{tot}/cm^{-3})=-1$ to explore the case of the recently-observed "hybrid" bursts, displaying a long-GRB-like radiative emission but happening outside star-forming regions and, thus, embedded in lower density gas (see e.g. \citealp{troja22}).

At all sampled times the ionisation degree of the ISM surrounding the GRB is highly structured and stratified, and it remains asymptotically stratified for at least years after the afterglow emission has completely faded (time dictated by the slow gas cooling and recombination rates, which both depend on the gas density). 
Close to the GRB all the elements are highly ionised, up to hydrogen-like and fully stripped. The degree of ionisation and the temperature decrease for increasing $r$ and approach the pre-burst values at a sufficient distance from the GRB.

However, for a given $N_H$ the relative abundances of high versus low ionisation ions strongly increases with $n_e^{tot}$, since the denser the medium, the smaller (and closer to the GRB) the radial interval enclosing that $N_H$. This can be seen by expressing the geometric dilution of the flux as a function of $N_H$, i.e. $F_{ion}(N_H) = F_{ion}(r_{in}) \cdot (r_{in}/r_{N_H})^2 \propto F_{ion}(r_{in}) \cdot (n_e^{tot}/N_H)^2$, i.e. at a given $N_H$ (enclosed in a radius $r_{N_H}$), the ionising flux is $\propto (n_e^{tot})^2$ and so less diluted for increasing density. As a result, the transmitted spectra will be different and, thus, high-resolution X-ray absorption spectra of GRB afterglows taken at different times after the explosions can yield a direct tomography of the GRB environment, as we will show in \S \ref{GRBspectra} below. 

\begin{figure}
\centering
\includegraphics[width=0.9\columnwidth]{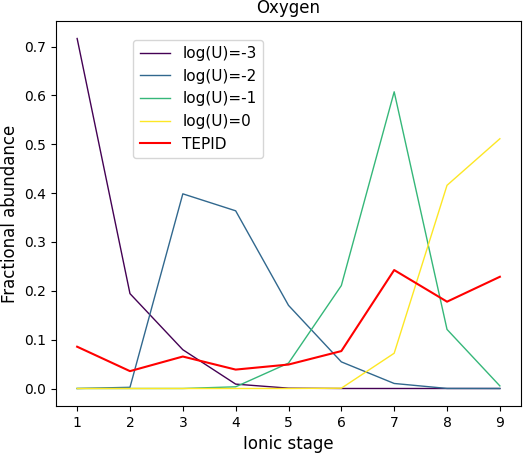}
\caption{Red line: relative abundances of the Oxygen ion stages for $log(n_e^{tot}/cm^{-3})=1, t=10^4 s$ averaged up to $log(N_H/cm^{-2})=21$. Blue to green lines: photoionisation equilibrium abundances for $log(U)$ from -3 to 0, averaged up to $log(N_H/cm^{-2})=21$.}
\label{eq-comp}
\end{figure}

\subsection{Absorption spectra}
\label{GRBspectra}
Fig. \ref{theor_spectra_nh}, left panel, shows the broad-band 0.1-10 keV absorption spectra imprinted on an incident GRB afterglow powerlaw continuum with photon index $\Gamma=2$ emerging from the first 100 pc of circum-burst medium, at $t=10^4$ s and for increasing $log(n_e^{tot}/cm^{-3})$, from -1 to 4 (and, thus, increasing $log(N_H/cm^{-2})$, from 19.5 to 24.5). The right panel shows a zoom-in on the 0.5 - 0.7 keV range, where oxygen lines from neutral (OI) to hydrogen-like (O VIII) transitions lie. Higher $n_e^{tot}$ probes larger and, on average, less ionised columns of circum-burst medium (see Fig. \ref{GRB_movie}). This implies emerging spectra attenuated by relatively high fractions of neutral (and so more opaque) gas for $log(n_e^{tot}/cm^{-3})=4$ and by increasingly larger fractions of ionised gas as the density decreases. 

Fig. \ref{theor_spectra_n} shows the same spectra for the same range of $log(n_e^{tot}/cm^{-3})$ and a fixed column density of $log(N_H/cm^{-2})=22$. This translates in a broad range of radii according to the electron density. For $log(n_e^{tot}/cm^{-3})=$4(=3), the radius is only $3.2(=0.32) pc$, therefore the medium is totally ionised and the spectra are unabsorbed (except for some lines for $log(n_e^{tot}/cm^{-3})=3$). On the other hand, for $log(n_e^{tot}/cm^{-3})=-1$ the radius is 32 kpc and the absorbing gas is mostly neutral.

The dependence on $n_e^{tot}$ is a peculiar feature of non-equilibrium photoionisation which cannot be reproduced with equilibrium, steady-state photoionisation models. As shown above, the simultaneous presence of all these ionisation stages is due to the stratification of the ionisation structure of the intervening medium, whose closer-in ionised regions have no time to recombine before the afterglow fades while farther-out zones receive a progressively geometrically diluted number of ionised photons and never reach high degrees of ionisation. 
The out-of-equilibrium ratio between different ionisation stages of a given element in the observed spectra is thus unique to the specific physical properties of the medium surrounding the GRB (namely its density profile and extent of the star-forming region in the line of sight direction) and is very different from what would be expected if the gas were in photoionisation equilibrium (see also \citealp{lazzati03}). 
This can be clearly seen in Fig. \ref{eq-comp}, where we compare the oxygen ion abundances predicted by TEPID for $log(n_e^{tot}/cm^{-3})=1$ and $t=10^4 s$ with those expected if the medium were in photoionisation equilibrium with four different values of the ionisation parameter. Red line shows the OI-IX abundances averaged over a gas column $log(N_H/cm^{-2})=21$ (i.e. 32 pc), while blue to green lines correspond to the equilibrium abundances (computed with Cloudy) for $log(U)$ from -3 to 0 and averaged over $log(N_H/cm^{-2})=21$. When the gas is in photoionisation equilibrium with the ionising source, ion abundances peak around a single ionisation stage (see also \citealp{kallman04}). Our time-evolving abundances, instead, clearly show the stratification of the medium.
Indeed, at equilibrium the abundance of each ionic stage is simply proportional to the ratio between its recombination and photoionisation rates. Here, instead, recombination is negligible with respect to ionisation (see Fig.\ref{rates-compare} in Appendix \ref{rates}), both because of the high ionising flux and of the relatively low values of $n_e$, and so the (rising) gas ionisation for increasing time is simply given by the continuous photoionisation of the gas due to the GRB radiative output.

The strong dependence of the afterglow transmitted spectrum on $n_e^{tot}$ offers a valuable opportunity to directly constrain it in observed spectra, by fitting afterglow spectra with time-evolving photoionisation models such as TEPID.

\subsection{Impact of the pre-burst conditions}
To quantify the impact of the initial ionisation of the surrounding medium we compare the results obtained under our assumption of an initial Wolf-Rayet like preionisation with those obtained by assuming an initially neutral medium. 
As shown in Fig. \ref{preion-initabunds}, stellar luminosity produces an ionisation of the surrounding medium up to far smaller distances than those covered by the GRB ionisation, especially for metals (i.e., elements heavier than He) which are the ones determining the gas spectroscopic appearance in the X-ray band. This is due both to the soft SED of stars, which can be described as blackbody radiation and, thus, is rather steeper than the typical GRB powerlaw SED, and to the much lower luminosity (around 18 orders of magnitudes for the physical settings of this paper). As a result, the impact of the initial temperature and abundances is negligible (see Appendix \ref{comp-preion} for all the details) and the absorption spectra are totally indistinguishable.

\section{Discussion}
\label{discuss}

\subsection{Current Limitations of the Code and Ongoing Upgrades}
Main limitation of the current version of TEPID is the assumption that all ions lie in their ground states at all times. 
In photo-ionised gas, levels above the ground with $\Delta E \sim kT$ get significantly and stably populated at densities larger than those for which collisional excitation rates are comparable to spontaneous de-excitation rates (e.g. \citealp{netzer13,kallman21}). For permitted transitions this implies densities $n_e >> 10^{10}$ cm$^{-3}$, at which proper time-evolving photoionisation of the intervening gas becomes less and less critical. However, we note that time-evolving effects at such high densities could manifest when the timescale of the luminosity variability is very short, as in the case e.g. of X-ray lags and Quasi-Periodic Oscillations in AGNs and compact sources (see e.g. \citealp{kara16,wang22}). We refer to \cite{garcia13} for further discussion on this point.

The lack of a time evolving electronic level population treatment prevents the calculation of diffuse and reflected line emission (and so an accurate radiative transfer treatment and the prediction of the gas emission spectrum) and also affects the correct ion balance computation and thus, in turn, the heating-cooling balance. This explains most of the differences seen in Fig. \ref{UT} between the equilibrium temperature computed with TEPID and those derived from Cloudy and XSTAR at $log(U) \simeq 1-3$, together with further improvement required to handle near-neutral gas states.

Our approximated radiative transfer considers outwards photoelectric absorption of the incident continuum throughout the gas cloud. Both the gas diffuse and reflected continua are not computed. They are only important when the diffuse and reflected radiation fields contribute significantly to the total emitted radiation. 
In the GRB scenario the incident ionising continuum largely dominates, as the ISM gets quickly photoionised by the incoming photons during the prompt and the early afterglow phases (i.e. from hundreds to thousands seconds, see Fig. \ref{GRB_movie}), while recombination continuum within the circum-burst medium is produced over much longer time-scales even in the densest ISM and star-forming regions.

\begin{figure}
\centering
\includegraphics[width=\columnwidth]{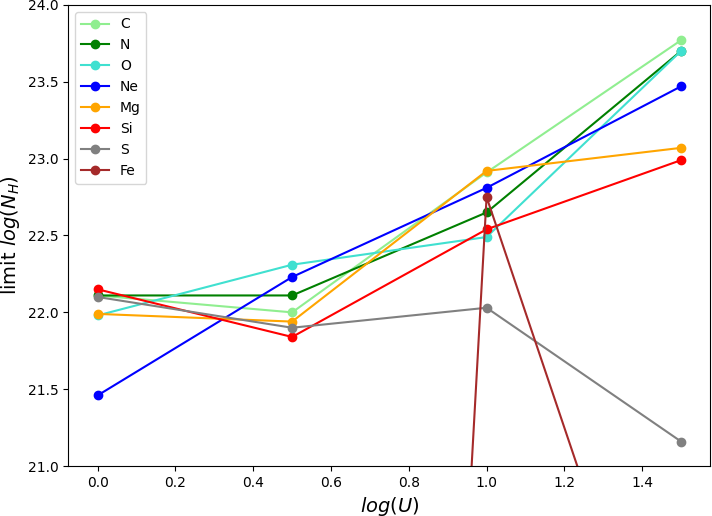}
\includegraphics[width=\columnwidth]{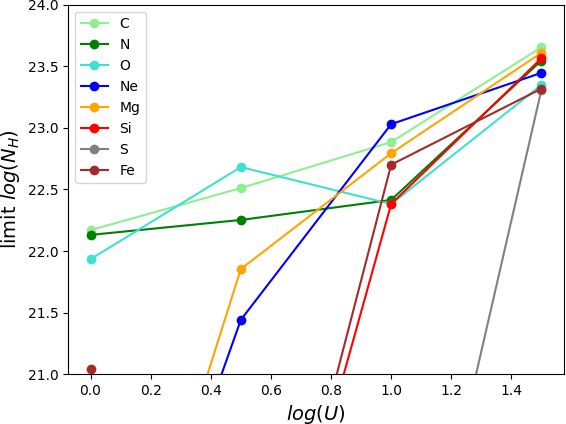}
\caption{Top: Maximum value of $N_H$ up to which the agreement between TEPID and Cloudy is within 50\% as a function of $log(U)$. Colour coding corresponds to different metals, see legend. Bottom: Same comparison of above, but between Cloudy- and XSTAR-predicted abundances.}
\label{logU-logNH}
\end{figure}

To assess the impact of the lack of the diffuse spectrum on the ionisation balance in the AGN scenario, we run TEPID at equilibrium (as discussed in \S \ref{phot-eq}) and compare the resulting ion abundances as a function of $N_H$ with those predicted by Cloudy in the range $0<log(U)<1.5$. Fig \ref{logU-logNH}, top panel, shows, for each metal included in TEPID and as a function of $log(U)$, the maximum $N_H$ up to which the agreement between the ion abundances computed with TEPID and Cloudy is within 50\% . To perform a meaningful comparison, for each value of $log(U)$ we limit to the most abundant ions, i.e. those with abundance $\geq 0.1$.
As expected the agreement improves for increasing ionisation, since the gas opacity, and thus its emission, decreases. The mismatch between Cloudy- and TEPID-computed abundances for Fe and S, which is significantly higher than all the other metals, is due to the lack of updated recombination rates for low ionisation levels (from Fe 1 to 11 and S 1), for which Cloudy uses custom-computed means while we employ the latest available rates from \citealp{m98}.
To give an idea of the scatter between different photoionisation equilibrium codes, bottom panel of Fig \ref{logU-logNH} reports the limit $N_H$ obtained by comparing Cloudy and XSTAR ion abundances. It can be seen the agreement is roughly of the same order as that between TEPID and Cloudy.

It must also be noted that we neglect photon scattering, which will lead to a sizeable underestimate of the gas opacity for Hydrogen-equivalent column densities above $1/\sigma_T = 1.5 \cdot 10^{24} cm^{-2}$, where $\sigma_T$ is Thompson cross section. See Appendix \ref{thompson sc} for further discussion on this point.

Finally, in the current version of TEPID we do not consider free-free absorption within the cloud. This is a significant source of heating at low ionisation degrees and temperatures and explains the cooler gas equilibrium temperatures predicted by TEPID at $log(U) \ls -1$ compared to  Cloudy and XSTAR (Fig. \ref{UT}). 

The next version of the code will self-consistently solve the level population equations and this, in turn, will allow us to implement a physically self-consistent time evolving radiative transfer, by following the approach outlined by \cite{vanadelsberg13}, and to compute the emerging absorption and emission spectra as a function of time, geometry and dynamics of the gas. 

We also stress that, for time-dependent effects to be detected in observed data, the ionised gas will have to spend a significant fraction of the observing time of each (time-resolved) spectral slice out of photoionisation equilibrium. The fulfilment of such condition strongly depends both on the observed source properties (variability amplitude and timescale of the ionising continuum, density and ionisation of the absorbing gas) and on the signal to noise of the observation and, thus, it requires an accurate evaluation on a case-by-case basis. 

However, we note that for GRB circumburst environments the gas recombination tends to be negligible with respect to ionisation (as discussed in \S \ref{GRBspectra} and Appendix \ref{rates-compare}) and, thus, the transmitted spectrum is mainly shaped by the progressive reduction of the gas opacity with time (see Fig. \ref{GRB_movie}). As we show in Appendix \ref{integrated spec}, this leads to absorption spectra which are dramatically different from those due to equilibrium absorbers even when integrated over the typical XMM-Newton observing times of GRB afterglows.

\subsection{Comparison with other Time-Evolving Photoionisation codes}
The upcoming advent of a new generation of X-ray spectrometers (see below) has increased the need for accurate modelling. Over the past few months two time-evolving ionisation codes similar to TEPID have been presented by \cite{sadaula22} and \cite{rogantini22}, optimised on the physics of ionised absorbers in AGNs. As TEPID, both codes solve ion and energy balances as a function of time, neglecting the calculation of diffuse and reflected continuum and line emission, and compute time-resolved absorbed X-ray spectra. The code of \cite{sadaula22} performs an approximated radiative transfer treatment similar to ours, while that of  \cite{rogantini22} considers only optically thin gas clouds. 

With respect to the above codes, TEPID is probably faster as it uses the in-flight adaptive temporal resolution algorithm described in \S \ref{adaptive time binning}. This reduces computing time at an affordable level both when the ionising source SED varies significantly over short timescales (e.g. variable AGNs and X-Ray Binaries, see Figs. \ref{tevol_abund}, \ref{tevol_teleNH}) and when the gas is geometrically thick, i.e. $\Delta r/r \gtrsim 1$ (see \S \ref{overview}), such as for the circum-burst medium around GRBs (see Figs. \ref{GRB_movie}). 

We compare TEPID and TPHO, the code of \cite{rogantini22}, by reproducing their "flaring lightcurve" case (see their \S 3.4), in which an optically thin gas, initially in photoionisation equilibrium, is exposed to a factor 10 increase of the incident ionising radiation within 20 kiloseconds. We obtain a remarkable agreement between the ionic abundances, temperature and energy balance as a function of time , as shown in detail in Appendix \ref{rogantini-comp}.

\subsection{Time-Resolved Spectral Analysis with Future X-ray Spectrometers}
The unprecedented energy resolution and sensitivity of the microcalorimeters onboard the next X-ray missions XRISM and Athena \citep{xrismWP,xifu22,athenaWP} will revolutionise our understanding of the gaseous environment of AGNs, compact sources and X-ray transients. Observations of gas undergoing photoionisation by variable sources will require appropriate time-evolving codes to be properly analysed and to meaningfully constrain the physical parameters of the gas. As shown in Fig. \ref{eq-comp}, time-evolving ionisation leads to unique ion abundance patterns which cannot be mimicked by any equilibrium code. Fitting single-epoch absorption spectra of a time-variable Warm Absorber with equilibrium models may yield to erroneous diagnostics (e.g. Nicastro et al., 1999, Krongold et al., 2007) and even suggest the presence of multiple gas layers with different ionisation parameters where only one non-equilibrium gas component is actually present \citep{rogantini22}. 

\section{Conclusions}
\label{conclus}
In this paper we present the new Time Evolving PhotoIonisation Device (TEPID). TEPID is a highly modular and flexible time-evolving photoionisation code that produces time-resolved absorption spectra of gas illuminated by variable ionising astrophysical sources. TEPID's time-resolved spectra can be stored in table models to be be used within spectral fitting packages (such as \textit{xspec}) to fit observed data. 
We apply TEPID to the case of AGN ionised absorbers and the circumbust medium around a GRB. Our main findings are:
\begin{itemize}
\item The equilibration time, $t_{eq}$ (Eq. \ref{teq}), is inversely proportional to the free electron density $n_e$ and dictates the timescale over which the gas readjust following a variation of the incident ionising luminosity. 
\item For ionising flux variations typical of AGNs and total (free + bound) electron densities $n_e^{tot}\gs 10^{10} cm^{-3}$ (i.e. Figs. \ref{tevol_ft} \ref{tevol_tele}, \ref{tevol_abund}), the absorbing gas reacts quickly and is close to photoionisation equilibrium at all times. In the highest density case, $n_e^{tot}=10^{12} cm^{-3}$, gas temperature and ion abundances are practically in instantaneous equilibrium with the ionising flux. For decreasing $n_e^{tot}$ (i.e., increasing $t_{eq}$), the gas is increasingly far from photoionisation equilibrium and its reaction to flux variations is both delayed and compressed. For the lowest explored density, $n_e^{tot}=10^4 cm^{-3}$, the gas is insensitive to flux variations.
\item The radial temperature and ionisation profiles of a star-forming region illuminated by a powerful GRB explosion strongly depend on the density of the circum-burst medium (see Figs. \ref{GRB_tele}, \ref{GRB_movie}). For all sampled densities,  $10^{-1} \leq n_e^{tot}/cm^{-3} \leq 10^4$, the resulting radial ionisation profile is highly structured and always far from what would be expected if the gas were instantaneously in photoionisation equilibrium with the ionising radiation field (see Fig. \ref{rates-compare}).
\item In both the AGN and the GRB scenarios the emerging X-ray absorption spectra are uniquely associated to the particular value of the gas electron density $n_e^{tot}$ (see Fig. \ref{AGN_spectra_t-n}, \ref{theor_spectra_nh}, \ref{theor_spectra_n}), which can therefore be tightly constrained through time-resolved spectroscopy. While preliminary and pioneering studies have been performed on low-resolution X-ray observations of such targets (e.g. N99, \citealp{kne07}, KP13), future high resolution and high-throughput X-ray spectrometers (e.g. the Resolve spectrometer of XRISM and the Athena X-IFU) will allow time-resolved spectroscopy to be systematically extended to sizeable samples of GRB afterglows and outflows from AGNs and X-Ray Binaries, enabling a detailed physical characterisation of the ionised gas under a number of different astrophysical scenarios (\citealp{xrismWP,athenaWP}; see \citealp{juranova22} for a feasibility study with the Athena X-IFU microcalorimeter).
\end{itemize}
Applications of TEPID to both GRB afterglow and AGN X-ray data will be presented in a series of follow-up papers. 

\begin{acknowledgements}
We thank the referee for their valuable comments, which significantly improved the paper. AL, FN, LP acknowledge support from the HORIZON-2020 grant “Integrated Activities for the High Energy Astrophysics Domain" (AHEAD-2020), G.A. 871158. We made use of the {\sc numpy} and {\sc matplotlib} packages \citep{numpy,matplotlib}.
\end{acknowledgements}

\bibliographystyle{aa}
\bibliography{sbs}

\begin{appendix}
\section{Radiation transfer}
\label{app-shield}

\begin{figure*}
\centering
\includegraphics[width=1.3\columnwidth]{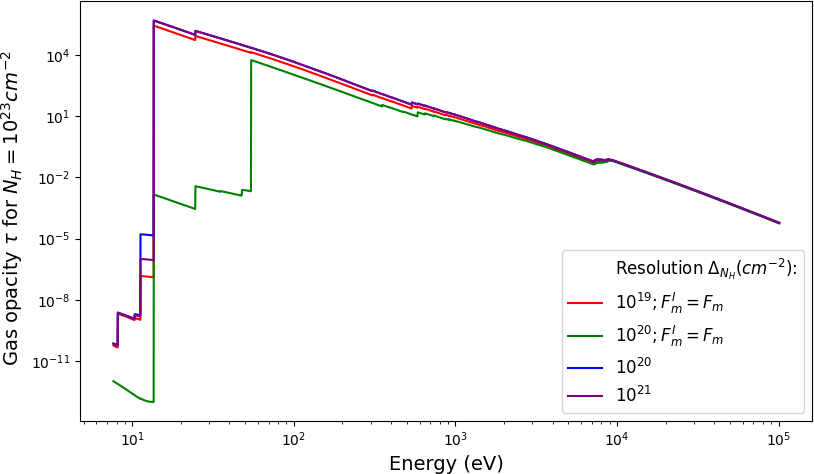}
\caption{Gas opacity $\tau$ as a function of energy for the circumburst environment of a GRB for $N_H=10^{23} cm^{-2}$. Green and red lines correspond to a spatial resolution $\Delta_{N_H}=10^{19},10^{20} cm^{-2}$, respectively, and do not account for the gas absorption within each shell. Blue and purple lines (superimposed throughout most of the energy range) correspond to resolutions of $\Delta_{N_H}=10^{20},10^{21} cm^{-2}$ and accounts for absorption following Eq. \ref{eq-shield}.}
\label{fig-shield}
\end{figure*}

Fig. \ref{fig-shield} shows the total gas opacity $\tau$ of the circumburst environment around a GRB with $log(n_e^{tot}/cm^{-3})=4, t=10^4 s$ and a total Hydrogen-equivalent column density $N_H=10^{23} cm^{-2}$. Red and green lines are computed ignoring the continuum absorption within each shell (i.e. using $F_m$ instead of $F'_m$, see Eq. \ref{eq-shield}) and setting a linear spatial resolution with steps of $\Delta_{N_H}=10^{19},10^{20} cm^{-2}$. In this setting, the shell ionisation is simply proportional to the incident spectrum ($F_m$) and, thus, it is overestimated, since the intra-shell absorption is not accounted for. This, in turn, leads to lower absorption when computing the transmitted spectrum $F_{m+1}=F_m \cdot e^{-\tau_m}$. This effect sums up throughout the total gas column, leading to a more ionised medium for thicker slabs: $\tau$ is indeed lower for higher $\Delta_{N_H}$.

Blue and purple lines, instead, are computed with $F'_m$ as in Eq. \ref{eq-shield} and linear $\Delta_{N_H}=10^{20},10^{21} cm^{-2}$, respectively. It can be seen that absorption is more accurately accounted for and, thus, the resulting $\tau$ is higher than in the previous cases. This demonstrates that Eq. \ref{eq-shield} allows to follow the propagation of the radiation with greater accuracy and, thus, allows to increase $\Delta_{N_H}$ up to two orders of magnitude (thus reducing considerably the computing time). Moreover, $\tau$ is practically identical for both values of $\Delta_{N_H}$, showing that a resolution of $10^{21} cm^{-2}$ is enough to properly sample the gas column.

\section{Impact of the preionisation on the GRB case}
\label{comp-preion}
\begin{figure*}
\centering
\includegraphics[width=1.6\columnwidth]{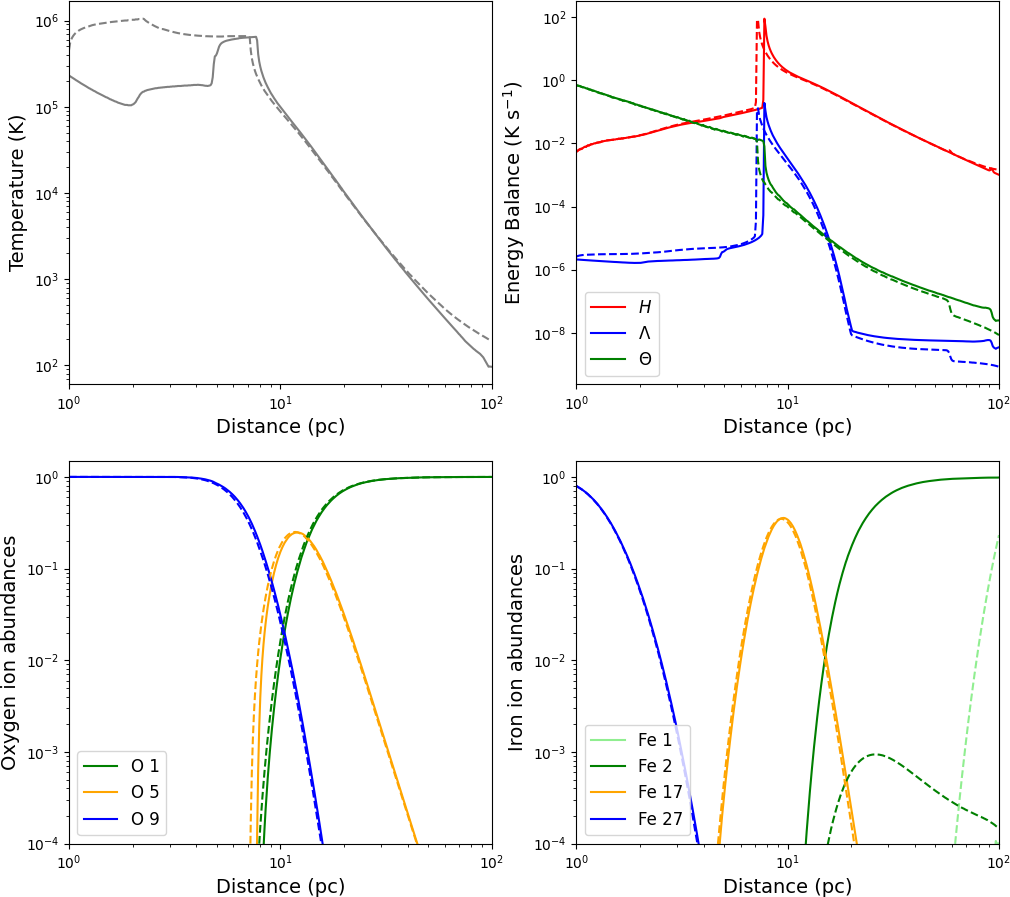}
\caption{Temperature (top left), energy balance (top right) and Oxygen, Iron ion abundances (bottom left and right) for $t=10^4 s, log(n_e^{tot})=2$ assuming both a Wolf-Rayet preionisation and an initially neutral medium (solid and dashed lines, respectively).}
\label{preion-comp}
\end{figure*}

Here we compare the gas temperature and ion abundances around the GRB obtained under two different assumptions concerning the initial ionisation of the gas: i) a Wolf-Rayet like ionisation, as described in \S \ref{phys-set} and adopted throughout the paper and ii) a totally neutral medium, with a typical ISM temperature of $100 K$ \citep{draine11}.
Figs. \ref{preion-comp} reports $T$, $H, \Lambda, \Theta$ and the ion abundances of O, Fe. Solid and dashed lines correspond to Wolf-Rayet preionisation and initially neutral medium, respectively. In all cases $t=10^4 s, log(n_e^{tot})=2$.
The differences between the two cases are practically negligible, since the huge GRB luminosity is able to ionise up to far higher distances than the area affected by the progenitor star and, therefore, the resulting ionisation is weakly dependent on the initial conditions.

\section{Oxygen ionic stages}
\label{rates}
\begin{figure*}
\centering
\includegraphics[width=1.8\columnwidth]{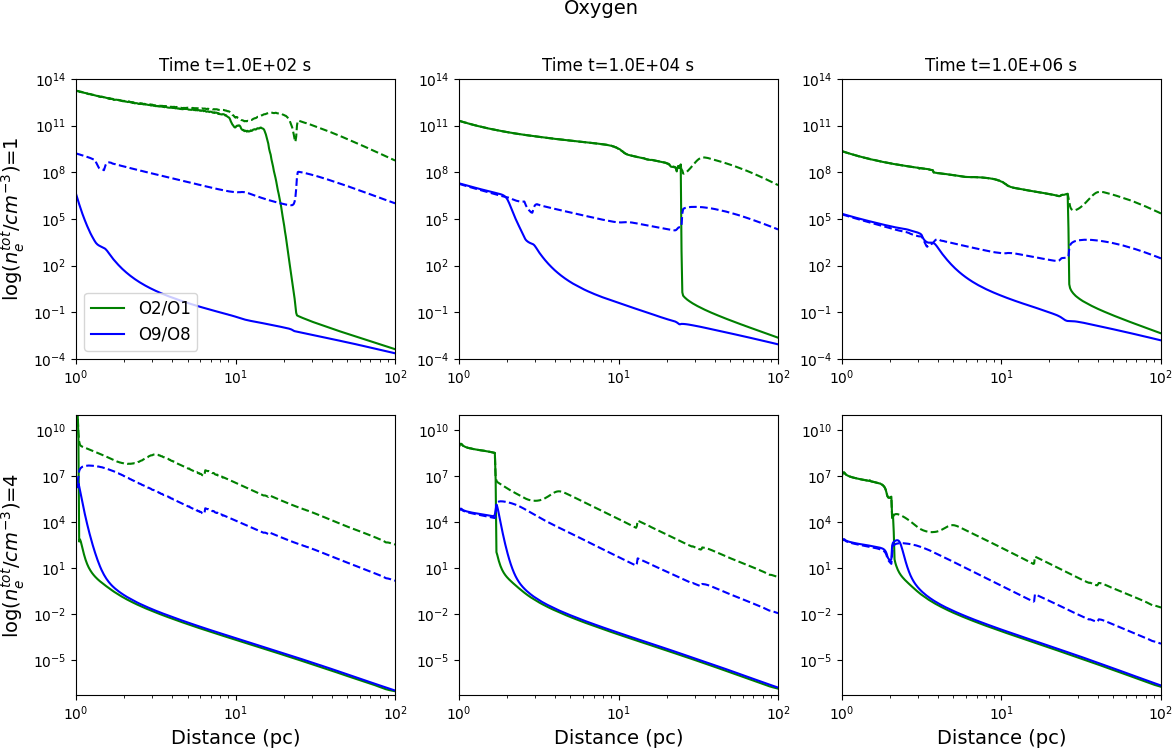}
\caption{Ratio between selected ionic stages of Oxygen (O IX / O VIII: blue solid line, O II / O I: green solid line). For comparison, dashed lines show the ratios expected if the gas were in photoionisation equilibrium. Top(bottom) row corresponds to $log(n_e^{tot}/cm^{-3})=1$(=4), while from left to right $t=10^2,10^4,10^6 s$.}
\label{rates-compare}
\end{figure*}

Fig. \ref{rates-compare} shows the ratio between the ion abundances of O IX and O VIII (blue solid lines) and between O II and O I (green solid lines). For comparison, we plot with dashed lines the values expected if the gas was in photoionisation equilibrium, which are given by the relation:
\begin{equation}
\frac{n_{X^{i+1}}}{n_{X^{i}}}=\frac{F_{X^i}}{\alpha_{rec}(X^i)}
\end{equation}
The ratios are in photoionisation equilibrium only within the innermost hot ionised region, where most of the oxygen is in the O IX and O VIII stage (Fig. \ref{GRB_movie}), while at higher distance the gas is under ionised and, thus, its ionisation will increases with time.

\section{Thompson scattering}
\label{thompson sc}
\begin{figure*}
\centering
\includegraphics[width=1.8\columnwidth]{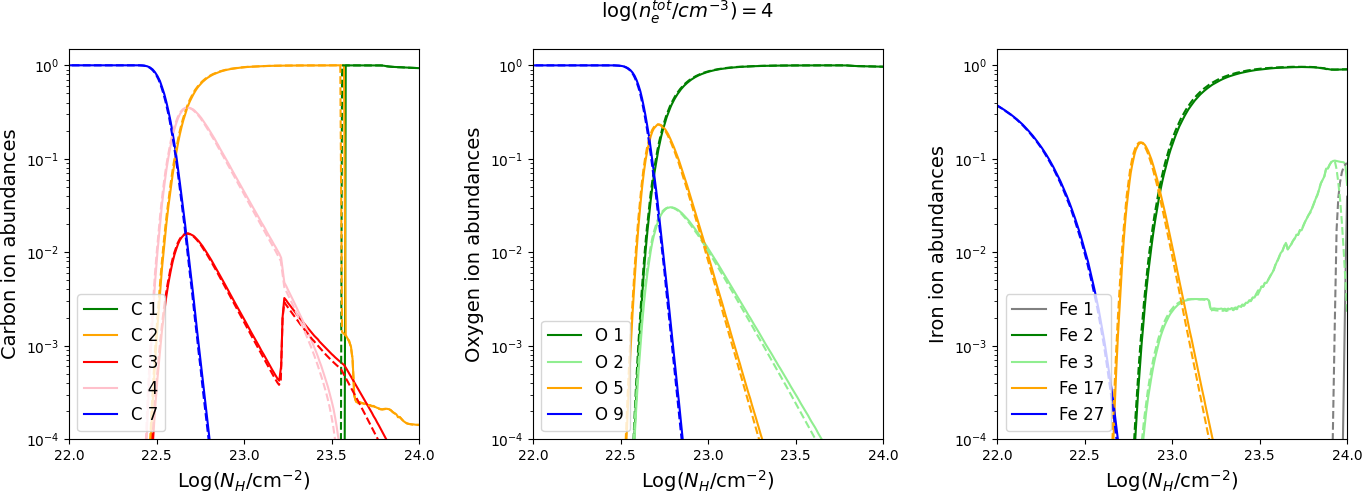} \\
\includegraphics[width=1.1\columnwidth]{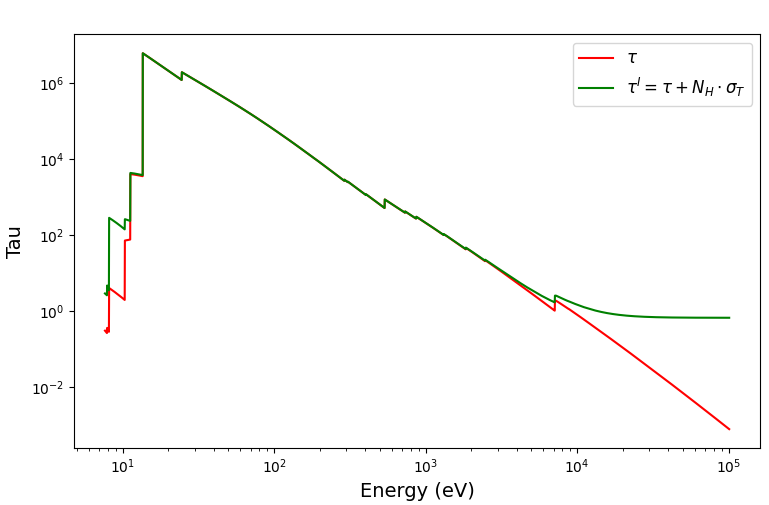}
\caption{\textit{Top:} From left to right: selected Carbon, Oxygen, Iron ion abundances (see legends) as a function of $N_H$, computed using the standard opacity $\tau$ and the Thompson-enhanced $\tau'$ (solid and dotted lines, respectively). \textit{Bottom:} Gas opacity for $N_H=10^{24} cm^{-2}$ as a function of the energy using $\tau$ and $\tau'$ (red and green line, respectively).}
\label{sigmaT}
\end{figure*}

The main effect of the electron scattering on our computations will be to deviate photons away from the line of sight and deplete the transmitted spectrum. To model this, we run a simulation in which we increase the gas opacity by using a modified opacity $\tau’(\nu) = \tau(\nu) + N_H \cdot \sigma_T$ , where $\sigma_T$ is the Thompson cross section and $\tau (\nu)$ is the opacity defined in Eq. \ref{tau_eq}. This framework, which roughly mimics the “open geometry” setting of equilibrium photoionisation codes such as Cloudy and XSTAR, maximises the opacity due to electron scattering, since it neglects photons scattered from different lines of sight towards our one. We find very small variations in both the ion abundances and the overall gas opacity. Figure \ref{sigmaT}, top panel, compares the ion abundances in the circumburst environment of a GRB obtained using $\tau$ (i.e., ignoring Thompson scattering, solid lines) and $\tau’$ (dotted lines). It can be seen that the differences are very small and mostly concentrated at $N_H>5 \cdot 10^{23} cm^{-2}$. In all cases, $n_e=10^4 cm^{-3}, t=10^4 s$ after the burst.

Figure \ref{sigmaT}, bottom panel, shows the gas opacity for $N_H=10^{24} cm^{-2}$ computed using $\tau$ and $\tau'$ (red and green lines, respectively). The two opacities overlap in most of the energy band. At low energy (E=10-20 eV) there is a small difference due to the slightly lower gas ionisation in the second case (see above panel). At $E \gtrsim 10$ keV, instead, the difference is due to floor given by Thompson opacity, whose value is $N_H\cdot\sigma_T = 10^{24} \cdot 6.65/10^{25}=0.6$.

\section{GRB afterglow spectra}
\label{integrated spec}

Figure \ref{TEPID-comp-spec} below shows the time-integrated TEPID spectrum (in red) of a GRB environment with $log(N_H/cm^{-2})=21, log(n/cm^{-3})=1$ between t=20 and 80 ks, corresponding to a typical XMM-Newton pointing of a GRB afterglow (A. Thakur et al., in prep.). For comparison, green lines correspond to photoionisation equilibrium solutions computed with the standard version of PHASE \citep{krongold03}, for $log(U)$ between -2 and 0 (colour coding, see legend). In all cases, the underlying continuum is a powerlaw with photon index $\Gamma=2$ and same normalisation.

It can be seen that also when accounting for the (finite) duration of a typical X-ray observation, the time-evolving ionisation leads to a unique absorption spectrum that cannot be mimicked by the equilibrium solutions, in quite good agreement with what shown in Fig. \ref{eq-comp} for a given instant of time (which would correspond to an “instantaneous” observation with a zero seconds duration).

\begin{figure*}
\centering
\includegraphics[width=1.6\columnwidth]{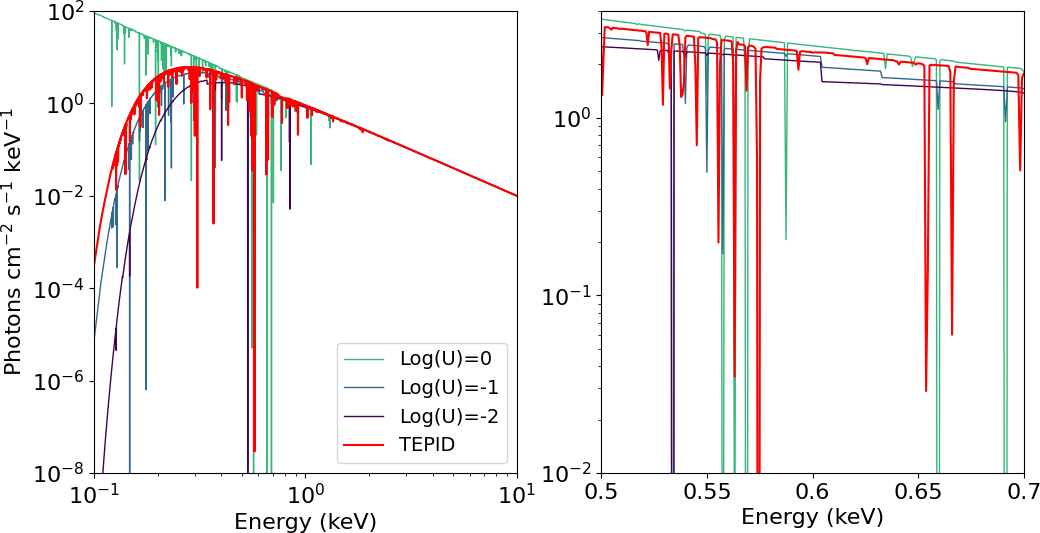}
\caption{Red: time-averaged spectrum for $log(N_H/cm^{-2})=21, log(n/cm^{-3})=1$ between t=20 and t=80 ks. Green spectra correspond to equilibrium absorbers, with $log(N_H/cm^{-2})=21$ and increasing ionisation parameter $log(U)$, see legend. Left panel shows the entire 0.1 -10 kev X-ray band, while right panel is a zoom-in on the 0-5-0.7 keV range.}
\label{TEPID-comp-spec}
\end{figure*}

\section{Comparison with the TPHO code}
\label{rogantini-comp}

\begin{figure*}
\centering
\includegraphics[width=0.35\paperwidth]{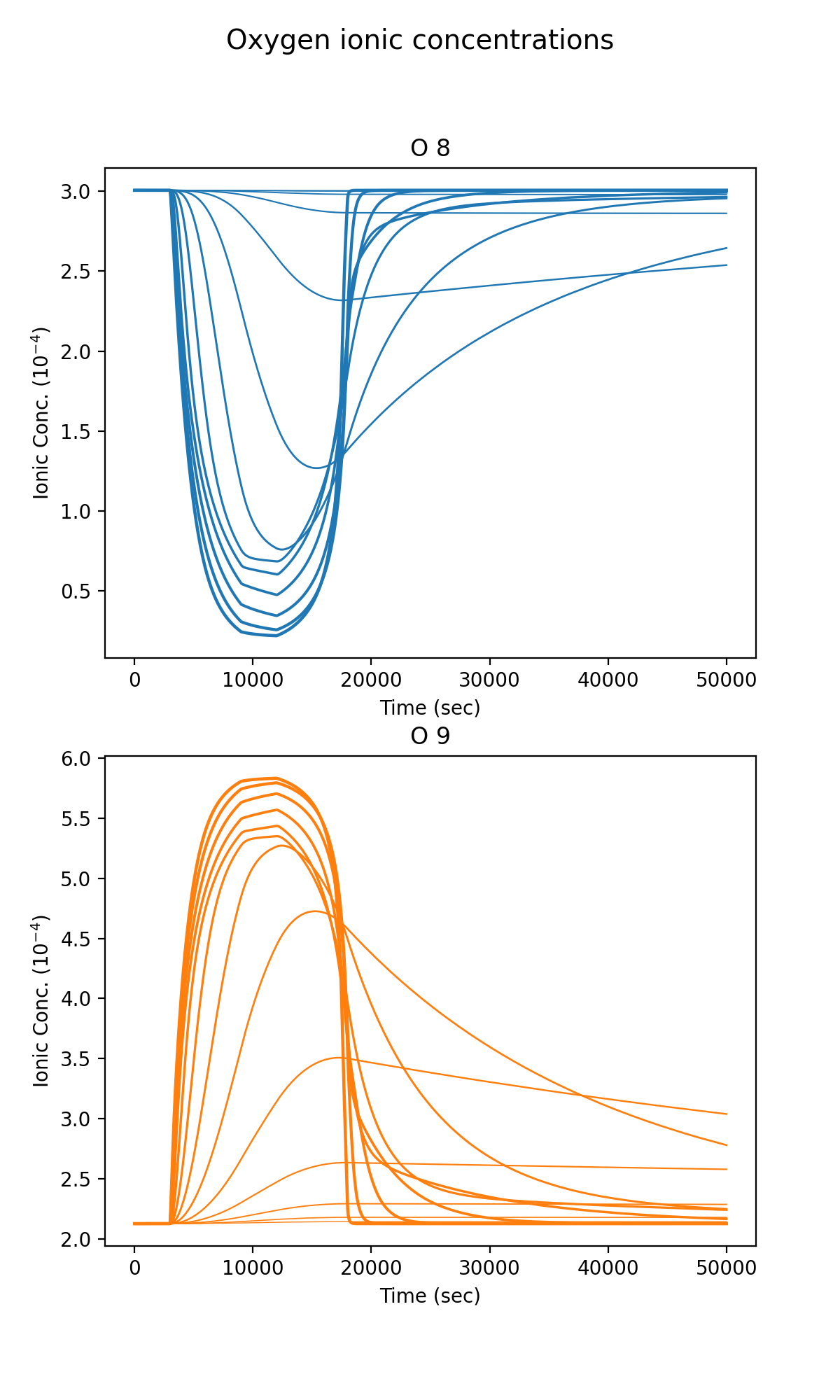}
\includegraphics[width=0.35\paperwidth]{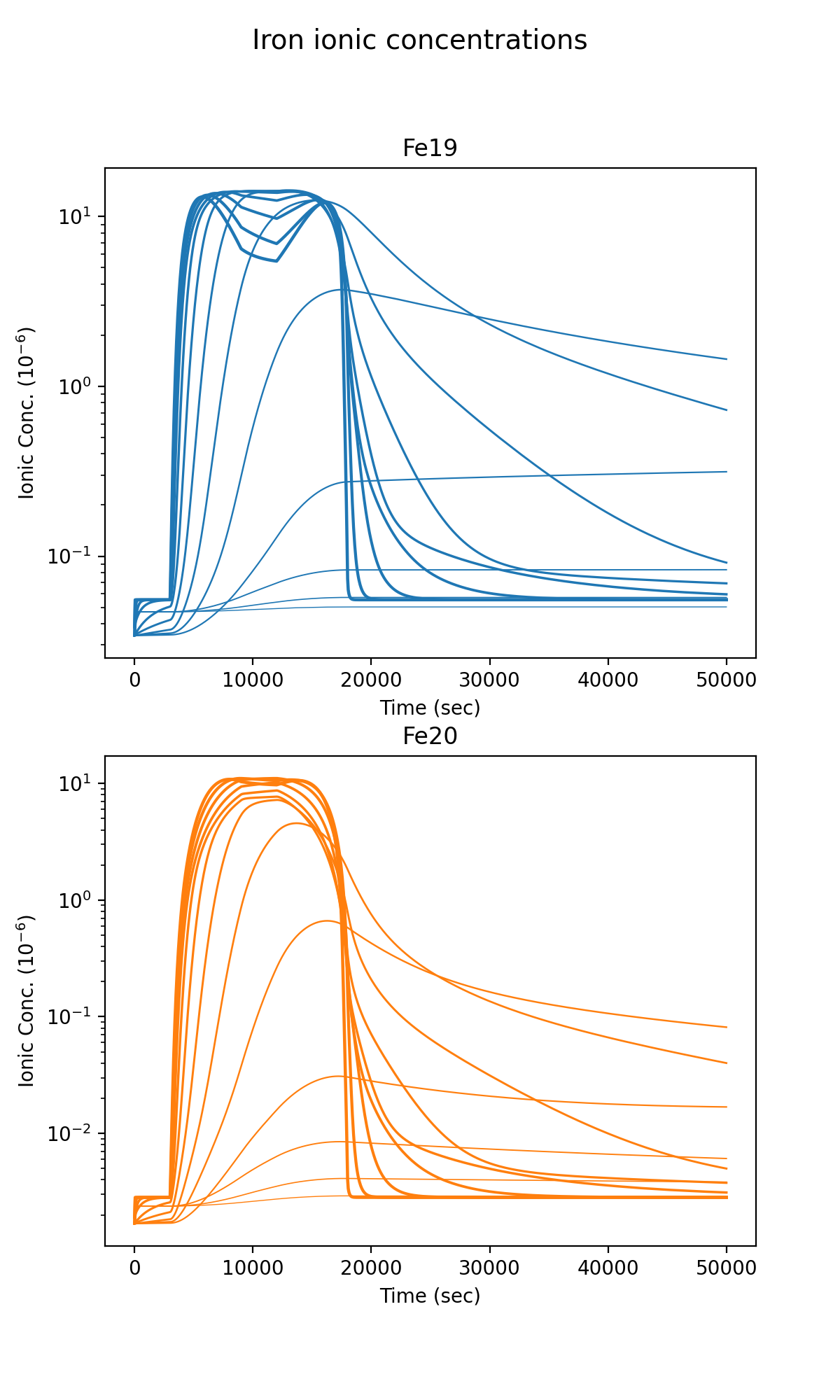}
\caption{Oxygen VIII, IX (left) and Iron XIX, XX (right) ionic concentrations as a function of time.}
\label{rogantini_ions}
\end{figure*}

\begin{figure*}
\centering
\includegraphics[width=0.75\paperwidth]{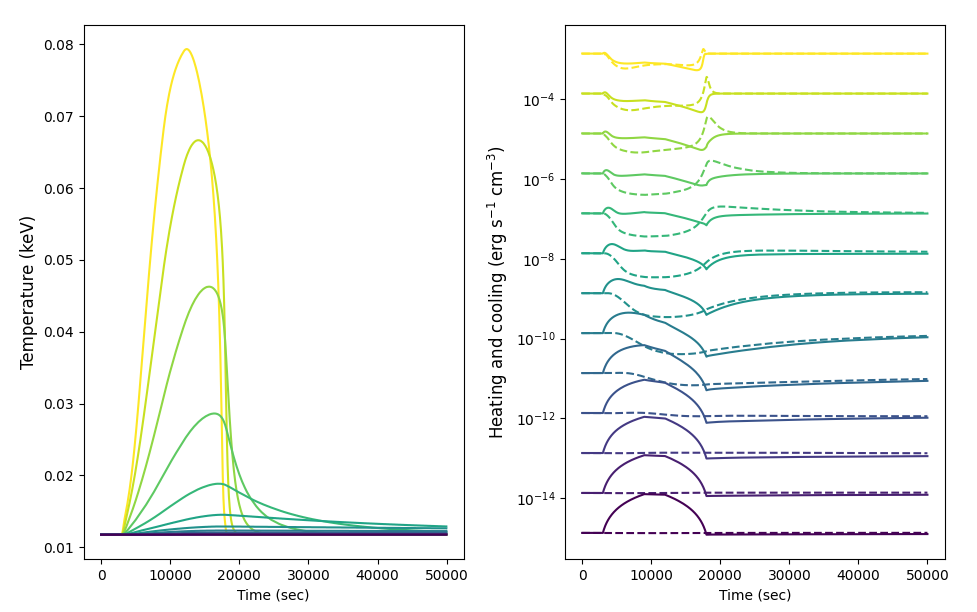}
\caption{Temperature (left) and heating and cooling rates (right, solid and dashed lines respectively).}
\label{rogantini_tele}
\end{figure*}

Figure \ref{rogantini_ions} show the ionic concentrations of Oxygen VIII, IX (left) and Iron XIX, XX (right), to be compared with those from TPHO reported in Fig. 7 in \cite{rogantini22}. It must be noted that ionic concentrations are defined in their paper as the product between the ionic abundance and the abundance of that element relative to Hydrogen. Thinnest to thickest lines correspond to increasing number density, from $log(n_e^{tot}/cm^{-3})=4$ to 10 with 0.5 step. Figure \ref{rogantini_tele} shows the temperature (top) and heating and cooling rates (bottom) for the same density range (darkest to lightest line), to be compared with their Fig. 8.

Notwithstanding the different atomic libraries (TEPID is based on the Cloudy database, while TPHO on the SPEX one) and the different structure of the codes, we obtain a remarkable agreement between the ionic abundances, temperature and heating and cooling rates as a function of time.

\end{appendix}
\end{document}